\documentclass{jfm}
\pdfoutput=1
\usepackage{hyperref}
\hypersetup{
	colorlinks=true,
	linkcolor=black,
	citecolor=black
}

\usepackage{graphicx}
\usepackage{natbib}
\usepackage{amssymb}
\usepackage{amsmath}
\usepackage{dashrule}
\usepackage{color}
\usepackage{psfrag}
\usepackage{bm}

\usepackage{calrsfs}
\usepackage{multirow,bigdelim}
\usepackage{mathtools}
\usepackage{layouts}

\newcommand\Ray{\mbox{\textit{Ra}}}  	
\newcommand\Nus{\mbox{\textit{Nu}}}  	
\newcommand\Web{\mbox{\textit{We}}}  	

\newcommand\ie{i.e.\ }
\newcommand\cf{cf.\ }

\newcommand{\romd}{\mathrm{d}}

\newcommand{\cs}[1]{{{\hypersetup{linkcolor=black,citecolor=black}{\color{black} #1}}}}
\newcommand{\dl}[1]{{{\hypersetup{linkcolor=black,citecolor=black}{\color{black} #1}}}}

\shorttitle{Transport mechanisms in VC with light droplets}
\shortauthor{C. S. Ng and others}
\title{Non-universal transport mechanisms in vertical natural convection with \\dispersed light droplets}
\author
{Chong Shen Ng$^1$\thanks{Email address for correspondence: c.s.ng@utwente.nl and}, Vamsi Spandan$^2$, Roberto Verzicco$^{1,3,4}$ and \\Detlef Lohse$^{1,5}$\thanks{ d.lohse@utwente.nl}
}

\affiliation{$^1$Physics of Fluids Group, Max Planck Center for Complex Fluid Dynamics, J.\,M.\,Burgers Center for Fluid Dynamics and MESA+ Research Institute, Department of Science and Technology, University of Twente, 7500AE Enschede, The Netherlands\\[\affilskip]
	$^2$School of Engineering and Applied Sciences, Harvard University, Cambridge, MA 02138, USA\\[\affilskip]
	$^3$Gran Sasso Science Institute - Viale F. Crispi, 7 67100 L'Aquila, Italy\\[\affilskip]
	$^4$Dipartimento di Ingegneria Industriale, University of Rome `Tor Vergata', Roma 00133, Italy\\[\affilskip]
	$^5$Max Planck Institute for Dynamics and Self-Organisation, 37077 G{\"o}ttingen, Germany
}

\date{?; revised ?; accepted ?. - To be entered by editorial office}

\begin{document}
\maketitle

\begin{abstract}

We present results on the effect of dispersed droplets in vertical natural convection (VC) using direct numerical simulations based on a two-way fully coupled Euler--Lagrange approach with a liquid phase and a dispersed droplets phase. For increasing thermal driving, characterised by the Rayleigh number, $\mbox{\textit{Ra}}$, of the two analysed droplet volume fractions, $\alpha = 5\times10^{-3}$ and $\alpha = 2\times 10^{-2}$, we find non-monotonic responses to the overall heat fluxes, characterised by the Nusselt number, $\Nus$. The $\Nus$ number is larger when the droplets are thermally coupled to the liquid. However, $\Nus$ retains the effective scaling exponents that are close to the ${1/4}$-laminar VC scaling, suggesting that the heat transport is still modulated by thermal boundary layers. Local analyses reveal the non-monotonic trends of local heat fluxes and wall-shear stresses: Whilst regions of high heat fluxes are correlated to increased wall-shear stresses, the spatio-temporal distribution and magnitude of the increase is non-universal, implying that the overall heat transport is obscured by competing mechanisms. Most crucially, we find that the transport mechanisms inherently depend on the dominance of droplet driving to thermal driving that can quantified by (i) the bubblance parameter $b$, which measures the ratio of energy produced by the dispersed phase and the energy of the background turbulence, and (ii) $\Ray_d/\Ray$, where $\Ray_d$ is the droplet Rayleigh number, which we introduce in this paper. When $b \lesssim O(10^{-1})$ and $\Ray_d/\Ray \lesssim O(100)$, the $\Nus$ scaling is expected to recover to the VC scaling without droplets, and comparison with $b$ and $\Ray_d/\Ray$ from our data supports this notion.

\end{abstract}

\section{Introduction}

Bubbles are ubiquitous. Within a liquid, they can play an important role in the transport of mass and heat. Such complex interactions of bubbles and liquids can be found in various applications and process technologies, for example in cooling systems of power plants, metallurgical industries, \dl{catalytic reactions} and in the mixing of chemicals \citep{Brennen.2005,Balachandar+Eaton.2010,Mathai+Lohse+Sun.2020}. One commonly studied class of bubble-liquid interaction is the bubble column \citep{Mudde.2005}, where liquid turbulence is generated and sustained by a rising swarm of bubbles. This form of turbulence is typically referred to as pseudo-turbulence
\citep{Lance+Bataille.1991,vanWijngaarden.1998,Mercado+Gomez+VanGils+Sun+Lohse.2010} or bubble-induced agitation \citep{Risso.2018}.

Various parameters can be controlled to modulate heat transport in a bubbly flow. For instance, one can use microbubbles to increase heat transport in the boundary layer \citep{Kitagawa+Murai.2013} or by inclining the domain \citep{Piedra+Lu+Tryggvason.2015}. The fluid properties can also be varied. For example, \cite{Deen+Kuipers.2013} studied the effects of bubble deformability and found localised increase of heat fluxes when bubble coalescence prevails in the near-wall region, whereas \cite{Dabiri+Tryggvason.2015} showed that nearly spherical bubbles tend to aggregate at the walls, which in turn agitate the thermal boundary layers and result in higher heat transport than for the case with deformable bubbles. From these studies, one key observation that can be made is that heat transport enhancement has been largely linked to boundary layer effects, \eg\,thinning of the thermal boundary layers or ejection of thermal plumes. On the other hand, a recent experimental campaign using a homogeneous bubble column found that the heat transport, characterised by the Nusselt number $\Nus$, not only increases by up to 20 times, but also becomes insensitive to the thermal driving of the background flow, characterised by the Rayleigh number $\Ray$ \citep{Gvozdic+Others.2018,Gvozdic+Others.2019}. The $\Ray$-insensitivity persists across a range of bubble volume fractions $\alpha$ between $5\times 10^{-3}$ and $5\times 10^{-2}$, implying that bubble-induced liquid agitation overwhelmingly dominates the heat transport mechanism across the thermal boundary layers. Indeed, the multifold enhancement in $\Nus$ is consistent with engineering estimates in the design of bubble column gas-liquid reactors \citep{Deckwer.1980}. 

Is there, however, any link between bubbly flows that directly influence the boundary layers versus bubble column experiments? And if any, are the boundary layers affected by \dl{the} dispersed phase in a universal manner when $\alpha > 0$? In this paper, we ask the question of how other parameters, specifically the density ratio of the dispersed phase to liquid phase, influence heat transport. Inspired by the water column experiments in \cite{Gvozdic+Others.2018} and to make contact with recent studies, we selected a setup of natural convection in a rectangular cell containing a dispersed phase consisting of freely rising and deformable light droplets.

\begin{figure}
	\centering
	\centerline{\includegraphics[trim=0 -1pc 0 -1pc,clip=true,scale=1]{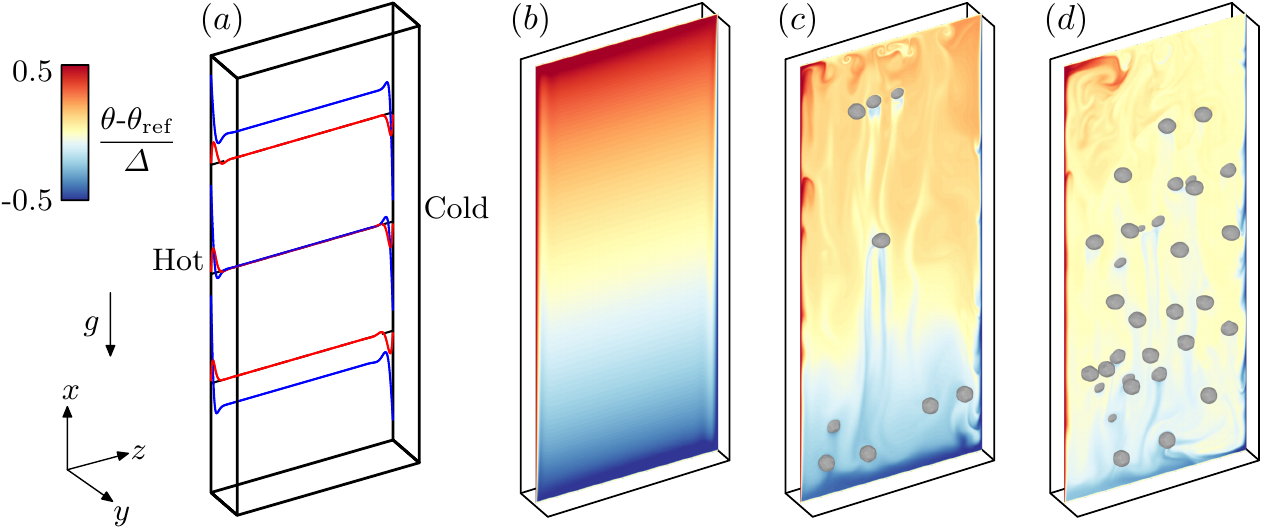}}
	\caption{\label{fig:setupvc}Visualisations of instantaneous temperature fields for VC at Rayleigh number of $2\times 10^8$. In ($a$), the red and blue curves correspond to mean velocity and temperature profiles at $x = 0.25L_x$, $0.5L_x$ and $0.75L_x$, respectively, at $\alpha = 0$. Volume fractions shown are ($b$) $\alpha = 0$, ($c$) $5\times 10^{-3}$, and ($d$) $2\times 10^{-2}$. Rendered flow fields are for droplets with mechanical coupling.}
\end{figure}
The model setup of the flow is thermal natural convection, in particular, a flow sustained by applying a temperature difference between two opposing walls. Classical examples of thermal natural convection include Rayleigh--B\'{e}nard convection \citep{Ahlers+others.2009}, where the hot wall is at the bottom and the cold wall at the top, and horizontal convection \citep{Hughes+Griffiths.2008,Shishkina+Grossmann+Lohse.2016}, where heating and cooling is applied at the same horizontal level. When the flow is confined between a hot vertical wall and a cold vertical wall, gravity acts orthogonal to the heat flux and this setup is referred to as vertical natural convection (VC). For confined VC, the bulk flow is quiescent (see mean profiles in figure \ref{fig:setupvc}$a$ and visualisation in \ref{fig:setupvc}$b$) and at low $\Ray$, the laminar-like boundary layers are expected to dominate heat and momentum transport \citep{Shishkina.2016.momentum}. This flow is unlike the unconfined, doubly-periodic VC \citep{Ng+Ooi+Lohse+Chung.2014,Ng+Ooi+Lohse+Chung.2017} where a mean shear is present and determines heat transport in the bulk flow region \citep{Ng+Ooi+Lohse+Chung.2018}. Hereinafter, we refer to the rectangular VC cell setup as VC, for simplicity.

When light droplets are introduced into VC, we ask two specific questions:
\begin{itemize}
	\item 	Are the heat and momentum transport statistics universal for droplets (\ie\,when the density of droplets are close to the density of the liquid)?
	\item 	How important is the role of thermal coupling between the droplets and the liquid?
\end{itemize}

To answer these questions, we perform direct numerical simulations (DNS) of VC with droplets where we have control over the density ratio and thermal coupling of the droplet phase to \dl{the} liquid phase. The droplets are fully coupled to the liquid phase DNS using the immersed boundary method (IBM) and the interaction potential approach, both of which are versatile numerical methodologies to simulate fully-coupled fluid flows with deformable interfaces \citep[\eg][]{Spandan+Verzicco+Lohse.2018,Meschini+Others.2018,Viola+Meschini+Verzicco.2020}. Furthermore, IBM offers some computational advantages over existing numerical methods for multiphase flows (\eg\,volume-of-fluid, level-set and front tracking), for instance, the underlying discretised grid is fixed and no sharp interfaces need to be resolved \citep{Spandan+Others.2017}. Recent advancements in the numerical methodology have allowed the use of sparser discretisations of the deformable interface relative to the underlying grid \citep{Spandan+Others.2018.fast} without compromising numerical accuracy, further easing the computational requirements for large-scale multiphase flows. \dl{The disadvantage of IBM, however, is that droplet coalescence or splitting is hard to model and correspondingly in this paper we refrain from attempting to do so.}

Our paper is organised as follows: in \S\,\ref{sec:FlowSetup}, we first describe the flow setup and numerical details for the fluid and dispersed phase. In \S\,\ref{sec:SimulationResults}, the numerical results are examined in detail. By analyzing the near-wall heat fluxes (\S\,\ref{sec:LocalNu}) and wall-shear stresses (\S\,\ref{sec:LocalCf}), we relate the droplet driving dynamics to changes in the near-wall statistics. In \S\,\ref{sec:ComparisonWithExp}, we discuss and compare the influence of our selected parameters and the experimental parameters as reported in \cite{Gvozdic+Others.2018}. Finally, in \S\,\ref{sec:ConclusionsOutlook}, we summarise our results and provide an outlook.

\section{Flow setup}\label{sec:FlowSetup}

Our reference setup is the single-phase VC flow (figure \ref{fig:setupvc}$b$), which is a buoyancy driven flow confined between two differentially heated vertical walls and two adiabatic horizontal walls. This reference flow will be referred \cs{to} as the liquid carrier phase. The flow is governed by mass conservation, balances of momentum and energy conservation which \dl{within} the Boussinesq approximation read,
\begin{align}
\partial_i u_i	&=	0, \label{eqn:ContEqn} \\
\partial_t u_i + u_j \partial_j u_i &=
-\dfrac{1}{\rho_{ref}}\partial_i p + \delta_{i1} g \beta (\theta-\theta_{\text{ref}}) + \nu\partial^2_j u_i + f_{i},	\label{eqn:MomEqn} \\
\partial_t \theta + u_j \partial_j \theta &= \kappa\partial^2_j \theta + q_i, \label{eqn:TempEqn}
\end{align}
where $\partial_t \equiv \partial/\partial t$, $\partial_i \equiv \partial/\partial x_i$, ($i,j=1,2,3$) and repeated indices imply summation. In \eqref{eqn:MomEqn}, $f_i$ is the back-reaction forces of the dispersed phase on the fluid and for single-phase VC, $f_i = 0$. The thermal analogue to $f_i$ in \eqref{eqn:MomEqn} is $q_i$ in \eqref{eqn:TempEqn}, which we selectively enable or disable in the present study. We define $\rho_{\text{ref}}$ as the reference density, $\theta_{\text{ref}}$ as the reference temperature, and $\beta$ is the thermal expansion coefficient of the fluid, $\nu$ the kinematic viscosity and $\kappa$ the thermal diffusivity, all assumed to be independent of temperature. The unit length is defined as the distance between the heated plates, $L_z$, and the streamwise and spanwise domain lengths are $L_x = 2.4L_z$ and $L_y = 0.25L_z$, respectively. Hereinafter, all lengthscales are non-dimensionalized by $L_z$. No-slip and no-penetration boundary conditions are imposed on the velocity at all four walls, whereas periodic boundary conditions are imposed in the $y$-direction. The left and right walls are imposed with temperatures hotter and cooler than the reference temperature $\theta_{\text{ref}}\equiv (\theta_h+\theta_c)/2$. The control parameters are the Rayleigh and Prandtl numbers, which are respectively defined as
\begin{subeqnarray}\label{eqn:DimensionlessNum}
	\gdef\thesubequation{\theequation \mbox{\textit{a}},\textit{b}}
	\Ray \equiv \dfrac{g\beta\Delta L_z^3}{\nu\kappa},\quad \Pran \equiv \dfrac{\nu}{\kappa},
\end{subeqnarray}
\returnthesubequation
where $\Delta \equiv \theta_h - \theta_c$. The aspect ratio can also be an additional control parameter for confined thermal convection problems \citep{vanderPoel+others.2011,Zwirner+Shishkina.2018}, but at present, we restrict our analyses to a fixed value. Our simulations cover the values of $\Ray = 1.3\times 10^8$-$1.3\times10^{9}$ and for $\Pran = 7$, corresponding to water.

The typical flow response is described by the Nusselt and Reynolds numbers,
\begin{subeqnarray}\label{eqn:DimensionlessNumResponse}
	\gdef\thesubequation{\theequation \mbox{\textit{a}},\textit{b}}
	\Nus\equiv \dfrac{f_wL_z}{\Delta \kappa},\quad \Rey\equiv \dfrac{U_sL_z}{\nu},
\end{subeqnarray}
\returnthesubequation
which quantify the dimensionless heat flux and degree of turbulence, respectively.  In (\ref{eqn:DimensionlessNumResponse}$a$), $f_w \equiv -\kappa (\romd\overline{\theta}/\romd z)|_w$ is the wall heat flux and $(\cdot)|_w$ denotes the wall value. $U_s$ is the `wind'-based velocity scale for VC \citep{Ng+Ooi+Lohse+Chung.2014} and accordingly, we set $U_s\equiv \overline{u}_{\text{max}}$, which is the maximum mean vertical velocity. The notation $\overline{(\cdot)}$ denotes time and $xy$-averaged quantities, and the notation $(\cdot)^\prime$ denotes the fluctuating component, \eg\,$u' = u - \overline{u}$. With the addition of the thermal forcing term, $q_i$, in \eqref{eqn:TempEqn}, a different definition for $\Nus$ becomes necessary because $(\romd \overline{\theta}/\romd z)|_{z=0} \neq (\romd \overline{\theta}/\romd z)|_{z=L_z}$ and the mean temperature equation now obeys $\overline{w'\theta'}(z) - \kappa \partial_z \overline{\theta} - \overline{q_i} = \text{const.}$. To overcome this difficulty, we employ the dissipation rate-based definition for the Nusselt number:
\begin{equation}
\Nus \equiv \dfrac{\varepsilon_\theta}{\kappa (\Delta/L_z)^2} = \dfrac{\langle \theta_{h} (\romd \overline{\theta}/\romd z)_{h} - \theta_{c}(\romd \overline{\theta}/\romd z)_{c} \rangle}{\Delta^2/L_z} + \dfrac{\langle \theta \cdot q_i \rangle}{\kappa (\Delta/L_z)^2}, \label{eqn:NuDissip}
\end{equation}
where $\varepsilon_\theta$ is the volume-averaged thermal dissipation due to turbulent fluctuations and $\langle \cdot \rangle$ denotes time- and volume-averaged quantities. When $q_i=0$, \eqref{eqn:NuDissip} equals to (\ref{eqn:DimensionlessNumResponse}$a$). The definition in \eqref{eqn:NuDissip} is also a direct analogue to the drag reduction calculations for multiphase Taylor--Couette flows \citep[\eg][]{Sugiyama+Calzavarini+Lohse.2008,Spandan+Verzicco+Lohse.2018}, making it convenient when comparing heat transport at matched $\Ray$ (discussed in \S\,\ref{subsec:NuReRa}). Throughout this paper, we will use \eqref{eqn:NuDissip} when reporting values of $\Nus$, unless defined otherwise.

The droplets are fully resolved using IBM for deformable interfaces and the interaction potential approach \citep{deTullio+Pascazio.2016,Spandan+Others.2017,Spandan+Others.2018.fast}. The simulations are also coupled in a so-called four-way manner, \ie\,the simulation is capable to handle droplet-fluid forcing, fluid-droplet forcing, droplet-droplet collisions and droplet-wall collisions. Our numerical methodology differs from point-particle-type simulations with heat transport \citep[\eg][]{Oresta+Verzicco+Lohse+Prosperetti.2009}: Since the droplets with diameter $D$ (at the point of injection) are significantly larger than the turbulent Kolmogorov length-scale $\eta$, we therefore fully resolve the inhomogeneous hydrodynamic forces acting at the droplet interface. To illustrate this point, we wish to stress that $D/\eta \approx 7$--$19$ in our simulations. Here, $\eta \equiv (\nu^3/\varepsilon)^{1/4}$, where $\varepsilon \equiv \nu \langle( \partial u_i/\partial x_j)^2\rangle$ is the volume-averaged turbulent kinetic energy dissipation rate. The key points of our IBM are summarised in \S\,\ref{subsec:numericdetails}.

\subsection{Numerical details} \label{subsec:numericdetails}
\begin{table}
	\def\drawline#1#2{\raise 2.5pt\vbox{\hrule width #1pt height #2pt}}
	\def\spacce#1{\hskip #1pt}
	\begin{center}
		\def~{\hphantom{0}}
		\begin{tabular}{r@{}l@{}cccccccc}
			&  	& 	$\Ray$ 	& $\alpha$ & $\Delta x^+$ & $\Delta y^+$ & $\Delta z_{w}^+$ & $\Delta z_{c}^+$ & $T_{s}/(L_z/U_\Delta)$ & $T_s/\langle t_{d}\rangle$ \\ 
			&  & $(\times 10^{9})$ & $ (\times 10^{-2}) $ & \multicolumn{6}{c}{{\color{white}\hbox{\drawline{60}{0.5}}}} \\[5pt]
			 &	&	$0.1$	&  ~-~ &	0.8	 & 0.8 & 0.3 & 1.1 &	400 & -  \\
			 &	&	$0.2$	&  ~-~ &	1.0	 & 1.0 & 0.3 & 1.4 &	570 & -  \\
			 &	&	$0.4$	&  ~-~ &	1.3	 & 1.3 & 0.4 & 1.7 &	570 & -  \\
			 &	&	$0.7$	&  ~-~ &	1.3	 & 1.3 & 0.3 & 1.8 &	480 & -  \\
			 &	&	$1.3$	&  ~-~ &	1.6  & 1.6 & 0.4 & 2.2 &	470 & -  \\[6pt]
			
		 	 \multirow{10}{1em}{\rotatebox{90}{Mech.\,coupling}} 			
		 	 &	\ldelim\{{10}{1em} 
		 	 	&	$0.1$	&  $0.5$ &	0.9  & 0.9  & 0.3 & 1.2 &	250 & 44 \\
			 &	&	$0.2$	&  $0.5$ &	1.1  & 1.1  & 0.3 & 1.5 &	190 & 24 \\
			 &	&	$0.4$	&  $0.5$ &	1.4	 & 1.3  & 0.4 & 1.9 &	230 & 22 \\
			 &	&	$0.7$	&  $0.5$ &	1.3	 & 1.3  & 0.4 & 1.9 &	320 & 22 \\
			 &	&	$1.3$	&  $0.5$ &	1.6	 & 1.6  & 0.4 & 2.3 &	400 & 21 \\[6pt]
			
			 &	&	$0.1$	&  $2.0$ &	0.9  & 0.9  & 0.3 & 1.3 &	230 & 40 \\
			 &	&	$0.2$	&  $2.0$ &	1.1  & 1.1  & 0.3 & 1.5 &	220 & 27 \\
			 &	&	$0.4$	&  $2.0$ &	1.4	 & 1.3  & 0.4 & 1.9 &	240 & 22 \\
			 &	&	$0.7$	&  $2.0$ &	1.4	 & 1.4  & 0.4 & 1.9 &	300 & 21 \\
			 &	&	$1.3$	&  $2.0$ &	1.7	 & 1.7  & 0.5 & 2.4 &	390 & 21 \\[6pt]
			
			 \multirow{10}{1em}{\rotatebox{90}{Mech.$+$therm.\,coupling}} 
			 &	\ldelim\{{10}{1em}
				&	$0.1$	&  $0.5$ &	0.9  & 0.9  & 0.3 & 1.2 &	220 & 39 \\
			 &	&	$0.2$	&  $0.5$ &	1.1  & 1.1  & 0.3 & 1.5 &	210 & 27 \\
			 &	&	$0.4$	&  $0.5$ &	1.4	 & 1.3  & 0.4 & 1.9 &	260 & 25 \\
			 &	&	$0.7$	&  $0.5$ &	1.3	 & 1.3  & 0.4 & 1.9 &	320 & 22 \\
			 &	&	$1.3$	&  $0.5$ &	1.6	 & 1.6  & 0.4 & 2.3 &	410 & 22 \\[6pt]
			
			 &	&	$0.1$	&  $2.0$ &	0.9  & 0.9  & 0.3 & 1.3 &	200 & 36 \\
			 &  &	$0.2$	&  $2.0$ &	1.1  & 1.1  & 0.3 & 1.5 &	200 & 26 \\
			 &	&	$0.4$	&  $2.0$ &	1.4	 & 1.3  & 0.4 & 1.9 &	230 & 22 \\
			 &	&	$0.7$	&  $2.0$ &	1.4	 & 1.4  & 0.4 & 1.9 &	290 & 20 \\
			 &	&	$1.3$	&  $2.0$ &	1.7	 & 1.7  & 0.5 & 2.4 &	400 & 21 \\
		\end{tabular}
		\caption{Summary of simulation parameters. The corresponding number of grid points are $(n_x,n_y,n_z) = (960,96,384)$ for $\Ray \leqslant 0.4\times 10^{9}$ and $(n_x,n_y,n_z) = (1200,120,480)$ for $\Ray \geqslant 0.7\times 10^{9}$. $T_{s}/(L_z/U_\Delta)$ and $T_s/\langle t_{d}\rangle$ are the total simulation sampling interval in terms of the free-fall velocity and droplet rise times, respectively.}
		\label{tab:SimParam}
	\end{center}
\end{table}

The liquid phase is solved using DNS by a staggered second-order accurate finite difference scheme and marched in time using a fractional-step approach \citep{Verzicco+Orlandi.1996}. We employ equal grid spacings in the $x$ and $y$ directions, whereas the $z$ direction is stretched using a Chebychev type clustering. The selected resolutions are constrained by considerations of three issues:
\begin{enumerate}[i]
	\item 	the resolution at the corner flow regions,
	\item 	the resolution at the bulk flow,
	\item 	minimum number of gridpoints per droplet diameter.
\end{enumerate}
\cs{Concerning} point (i), we based our estimate from the minimum resolution guidelines \cs{proposed for laminar-like thermal convection simulations \citep{Shishkina+Stevens+Grossmann+Lohse.2010}}. As a check, a coarser simulation with 20\% less grid points results in $\Nus$ values that are within 0.5\%, indicating good convergence for our resolution. For point (ii), we determined that max[$\Delta x^+_i \equiv \Delta x_i/\delta_\nu$]~$\approx 2.4$ (details in table \ref{tab:SimParam}), where $\Delta x_i$ are the grid spacings in each $i$th direction and $\delta_\nu \equiv \nu/u_\tau$ is the viscous length scale based on the shear velocity scale $u_\tau \equiv [\nu(\romd \overline{u}/\romd z)|_w]^{1/2}$. Point (iii) is closely related to point (ii): although the bulk resolutions are coarse, they are carefully selected such that $D/(\text{max}[\Delta x_i]) \gtrapprox 28$, comparable \dl{to} other immersed boundary studies in turbulent flow with finite-size particles \citep{Wang+Vanella+Balaras.2019}. Other numerical strategies are certainly possible, such as employing uniform grid spacings \citep{Lu+Fernandez+Tryggvason.2005} or by eliminating walls in the simulations \citep{Uhlmann+Chouippe.2017}, however, these strategies are either limited by the Reynolds numbers, or can be computationally costly. The resolutions employed here are therefore a careful compromise for our values of droplet rise Reynolds numbers,
\begin{equation}
\Rey_d \equiv U_dD/\nu, \label{eqn:DropletRe}
\end{equation}
where $U_d$ is the time-averaged vertical rise velocity of the droplet. Finally, to justify this point, we compare our IBM resolutions with the minimum resolution conditions for a flow over a rigid sphere \citep{Johnson+Patel.1999}. Given that our maximum droplet rise Reynolds number, max[$\Rey_d$]$\approx 220$ (see figure \ref{fig:DropPDF}), for an equivalent sphere Reynolds number, the dimensionless boundary layer thickness at its stagnation point is $\delta_{sp}/D\approx 1.13\Rey_d^{-1/2} \approx 0.08$ \citep{Schlichting2000boundary}. Our simulation resolution assures that at least two grid points reside within the droplet boundary layer. It may be tempting to treat this grid resolution as inadequate, however, we emphasize that this estimate is not only based on the extreme boundary layer criterion at the stagnation point, it is also based on the maximum $\Rey_d$ value and largest grid spacing in our setup. Our IBM resolution improves at lower $\Rey_d$ (\ie\,for thicker droplet boundary layers) and for finer near-wall grid spacings.

The dispersed phase is simulated with the IBM using a fast moving-least-squares algorithm \citep{Spandan+Others.2017,Spandan+Others.2018.fast}. Two volume fractions are simulated: $\alpha = 5\times10^{-3}$ and $2\times 10^{-2}$ (see table \ref{tab:SimParam}). In addition to the sampling intervals for fluid simulations, $T_{s}/(L_z/U_\Delta)$, where $U_\Delta \equiv (g\beta\Delta L_z)^{1/2}$ is the free-fall velocity, the number of droplet flow-through cycles, $T_s/\langle t_{d}\rangle$ where $\langle t_{d}\rangle$ is the time-averaged droplet rise time, is also an important parameter. At least 20 droplet rise intervals are recorded for each simulation. Droplets that rise close to the top of the domain are removed and re-injected randomly at the bottom of the domain. This procedure ensures a constant droplet volume fraction and a uniform droplet spatial distribution. In addition, the interfacial temperature is set as the mean temperature of the immersed fluid.

For the droplet boundary conditions, we assume that the droplets have negligible thermal inertia and are surfactant-laden. The first assumption implies a small droplet Biot number, \dl{defined by $\Bi \equiv hD/k$ where $h$ is the heat transfer coefficient and $k$ is the thermal conductivity of the droplet interface}, so that the internal droplet temperature can be approximated by a uniform temperature in accordance with the lumped-capacitance model \citep{Wang+Sierakowski+Prosperetti.2017}. Owing to deformation, individual droplet volumes can vary slightly throughout the simulation, but fluctuate about a constant reference volume--this is the underlying model of the interaction potential approach. Effectively, the droplet boundary conditions are no-slip and impermeable for velocity, with a homogeneous time-dependent temperature at the interface (The thermal boundary conditions are discussed in \S\,\ref{subsec:ModelThermalCouple}). Indeed, for physical systems with surface-active impurities, droplet interfacial dynamics may be closely approximated by a no-slip interface \citep{Jenny+Dusek+Bouchet.2004,Duineveld.1995}. These simplified boundary conditions also have the added benefit that they can be handled easily from a numerical point-of-view, and hence, are computationally efficient given the size of the flow problem.

To quantify the droplet deformability, we define the Weber number, $\Web \equiv \rho_{\text{ref}} U_\Delta^2 D/\sigma$, which yields the ratio of inertia to capillary forces, \dl{where} $\sigma$ is the surface tension. In our simulation strategy, $\sigma$ is not prescribed explicitly. Rather, an additional tuning step is performed to obtain a set of interaction potential constants such that $\Web \approx 3\times 10^{-2}$, in accordance with the tuning criteria described in \cite{Spandan+Others.2017}. It is emphasised that in order to simplify existing continuum models, this tuning step is a necessary and felicitous step in the implementation of our numerical model. After extensive precursor simulations and checks, we decided to simulate droplets at half the density of the fluid, \ie\,$\hat{\rho}\equiv \rho_d/\rho_{\text{ref}} = 0.5$, which is within the numerical stability limit for explicit IBM time integration schemes \citep{Schwarz+Kempe+Frohlich.2015}. Another reason why the explicit formulation is typically favoured over implicit (\ie\,strongly coupled) approaches for the fluid-structure interaction is also because of its computationally inexpensive nature. The detailed explanation of the methodology is, however, beyond the scope of this paper. For an in-depth discussion of the formulation, we refer readers to the paper of \cite{Spandan+Others.2017}.

\subsection{Model for thermally coupled droplets} \label{subsec:ModelThermalCouple}
For the lumped-capacitance model, two simplifying assumptions are made: (i) the droplets do not generate heat, and (ii) the internal temperature fields (and therefore interfacial temperature) of the droplets are uniform. Based on these assumptions, the interfacial droplet temperature is updated at every timestep according to 
\begin{equation}
\dfrac{\partial \theta_b}{\partial t} = \left\langle -\dfrac{\kappa}{V_d} \oint_{S_d} \dfrac{\partial \theta}{\partial \mathbf{n}}\cdot \mathbf{n}\,\partial S_d \right\rangle_{S_d}, \label{eqn:EnergyConservation}
\end{equation}
where $\theta_b$ is the mean interfacial droplet temperature, $V_d$ is the volume of the droplet, $S$ is the droplet surface area and $\mathbf{n}$ is the outwardly-directed unit normal. $\langle\cdot\rangle_S$ denotes the surface-averaged quantity. The droplet surface temperature is initialised as the mean surface temperature at its injected location. After injection, the droplets rise and deform \cs{with respect to} their original state (a sphere with diameter $D$), but do not significantly change in volume. Our model is therefore simpler than other numerical models with thermal coupling, for instance, studies that consider droplet growth \cs{at} the boiling limit \citep[\eg][]{Oresta+Verzicco+Lohse+Prosperetti.2009,Lakkaraju+Others.2011} or models that rely on droplets with a constant geometry \citep[\eg][]{Wang+Sierakowski+Prosperetti.2017}. Our code was extensively validated in a recent study for convection-dominated dissolution of droplets \citep{Chong+Li+Ng+Verzicco+Lohse.2019}.

\subsection{Derivation of the droplet Rayleigh number} \label{subsec:dropletRa}

In addition to the control parameters defined in \eqref{eqn:DimensionlessNum}, we introduce the droplet Rayleigh number, $\Ray_d$, to quantify the droplet driving. It is defined as
\begin{equation}
\Ray_d \equiv \dfrac{\alpha gL_z^3}{\hat{\rho}\nu\kappa}, \label{eqn:bubbleRa}
\end{equation}
\cs{which is conveniently derived from scaling arguments of the governing equations.}

Following the definition of IBM for deformable interfaces/fluid-structure interaction, the droplet interface is represented by a \cs{network} of nodes \cs{evolved} by the interaction potential model \citep{deTullio+Pascazio.2016,Spandan+Others.2017}. The equation of motion for each node, $e$, moving with velocity $\mathbf{u}_e$ is
\begin{equation}
\dfrac{d \mathbf{u}_e}{d t} = \mathbf{F}^{h} + \mathbf{F}^{g} + \mathbf{F}^{i}. \label{eqn:ElementEoMDless}
\end{equation}
In \eqref{eqn:ElementEoMDless}, the terms are made dimensionless with the unit length $L_z$ and free-fall velocity $U_\Delta$. The forces contributing to the right-hand-side of \eqref{eqn:ElementEoMDless} are the hydrodynamic \cs{loads} $\mathbf{F}^{h}$, buoyancy $\mathbf{F}^{g}$ and \cs{internal forces} $\mathbf{F}^{i}$, where
\begin{subeqnarray}\label{eqn:ElementEoMDlessRHS}
	\gdef\thesubequation{\theequation \mbox{\textit{a}},\textit{b}}
	\mathbf{F}^{h} = \dfrac{L_z}{\hat{\rho}V_eU_\Delta^2} \int_{S} \tau \cdot \mathbf{n}\,\romd S \quad \text{and} \quad
	\mathbf{F}^{g} \equiv \left(1-\dfrac{1}{\hat{\rho}}\right)\dfrac{L_z}{U_\Delta^2}\mathbf{g}.
\end{subeqnarray}
\returnthesubequation
In (\ref{eqn:ElementEoMDlessRHS}$a$), $V_e$ is the volume of the node, but lacks a physical definition because the definition of the thickness of a liquid-liquid interface is not straightforward. To overcome this, following \cite{Spandan+Others.2017}, we treat $V_e$ as a free parameter and fix $V_e=1$. The model parameters for the internal forces $\mathbf{F}^{i}$ are then correspondingly tuned. $\mathbf{F}^{i}$ represents the surface forces acting on the nodes of the discretised droplet surface and is based on the principle of minimising the potential energy of immersed interface. Under external hydrodynamic loads, the network of nodes deform and stores potential energy into the system. The potential energy is subsequently converted to surface forces by differentiating the potentials with respect to the displacements of each node. The details of the individual potentials are described in \cite{Spandan+Others.2017}.

We focus on the second term $\mathbf{F}^{g}$ in (\ref{eqn:ElementEoMDlessRHS}$b$). Since (\ref{eqn:ElementEoMDlessRHS}$b$) represents the contribution from an isolated droplet \cs{and we are interested to define a parameter for collective droplet effects, it would be reasonable} to include the volume fraction parameter, $\alpha$. Therefore, for $\hat{\rho} < 1$, we define,
\begin{equation}
F^{g}_\alpha \sim \dfrac{\alpha gL_z}{\hat{\rho}U_\Delta^2} \eqqcolon \dfrac{\Ray_d}{\Ray}, \label{eqn:RabubbleDivRa}
\end{equation}
which quantifies the relative dominance of droplet driving to thermal driving.

Other dimensionless parameters similar to $\Ray_d/\Ray$ have also been proposed for different flow configurations, but these require \textit{a priori} knowledge of the dispersed phase dynamics and/or flow statistics. For example, \cite{Climent+Magnaudet.1999} proposed the Rayleigh number expression, $\Ray_{CM} \equiv \rho g \alpha H^3/(\nu U_b)$ ($H$ is the height of the liquid layer and $U_b$ is the relative rise velocity of the bubble), to quantify bubble-induced convection. \dl{Based on the notion of pseudo-turbulence \citep{Lance+Bataille.1991}, which is defined as the fluctuating energy induced by the passage of bubbles under non-turbulent conditions, \cite{vanWijngaarden.1998}} proposed the so-called bubblance parameter $b \equiv (1/2)U_b^2\alpha/u_0^{2}$ ($u_0$ is the vertical velocity fluctuations of background turbulence). Since $\Ray_d/\Ray$ is a natural control parameter for VC with light droplets, we therefore use this ratio as input for our simulations. Note that $\Ray_d$ is constant for a given $\alpha$ and therefore $\Ray_d/\Ray$ reduces with increasing $\Ray$ (this is equivalent to an \textit{increase} in Froude number with increasing $\Ray$.). To make the simulations of the fluid-structure interaction tractable, we also run the simulations at $g/200$. The resulting $\Ray_d/\Ray$ is $5\times 10^{-4}$~--~$5\times 10^{-5}$ for $\alpha = 5\times10^{-3}$ and $2\times 10^{-3}$~--~$2\times 10^{-4}$ for $\alpha = 2\times10^{-2}$.

\section{Droplet influence on flow statistics and profiles} \label{sec:SimulationResults}
In this section, we analyse the results for $0 \leqslant \alpha \leqslant 2\times 10^{-2}$, starting with a discussion of the droplets statistics.

\subsection{Distribution of droplet aspect ratio versus bubble Reynolds number}
\begin{figure}
	\centerline{\includegraphics[trim=0 0 0 -1pc,clip=true,scale=1]{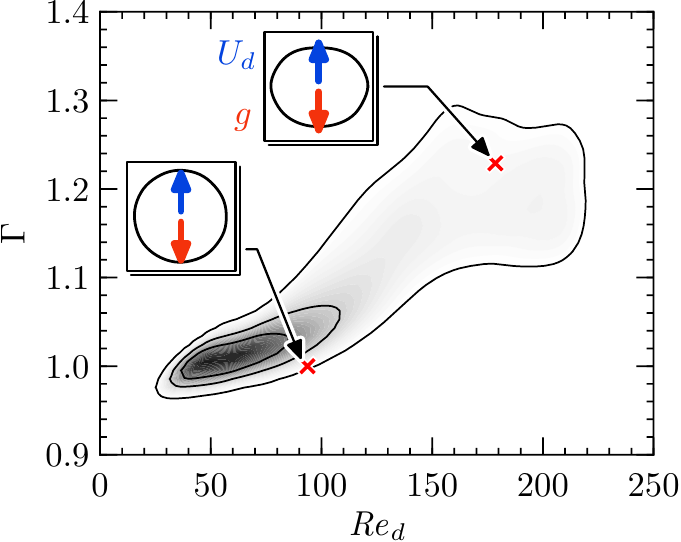}}
	\caption{\label{fig:DropPDF}Joint p.d.f.~of droplet deformations characterised by $\Gamma$ versus the droplet rise Reynolds number, $\Rey_d$, averaged over all cases. Outermost contour level is 0.03 and the contours are spaced 0.06 apart. Inset plots show representative two-dimensional droplet shapes for two $\Rey_d$ values. Also shown are the directions of the droplet rise velocity $U_d$ and gravity $g$.}
\end{figure}
From our simulations, the maximum \cs{droplet Reynolds number is} $\Rey_d \approx 220$ and \cs{its} time-averaged value is, $\langle \Rey_d\rangle \approx 100$. As the droplets rise, they undergo deformation from the interfacial hydrodynamic loads. In figure \ref{fig:DropPDF}, we characterise the deformation of the droplets in our simulations using the aspect ratio, $\Gamma$, of the major to minor axes, which are determined by fitting two-dimensional Fourier descriptors \citep{Duineveld.1995,Lunde+Perkins.1998} to the projected droplet outlines in the $xy$- and $xz$-plane. The joint probability density distribution in figure \ref{fig:DropPDF} shows that the droplets undergo moderate deformation between $\Gamma \approx 1$ to $\Gamma \approx 1.3$, agreeing with the relatively small $\Web$ values. Visual inspections of the instantaneous shapes (insets of figure \ref{fig:DropPDF}) show that the spherical droplet loses its fore-aft symmetry, with the front of the droplet becoming flatter than the back. Due to the relatively moderate $\Rey_d$ values, we do not observe droplet path instabilities throughout our simulations.

\subsection{Profiles of mean vertical velocity and temperature} \label{subsec:MeanProfiles}
\begin{figure}
	\centering
	\centerline{\includegraphics[trim=0 -0.5pc 0 -1pc,clip=true,scale=1]{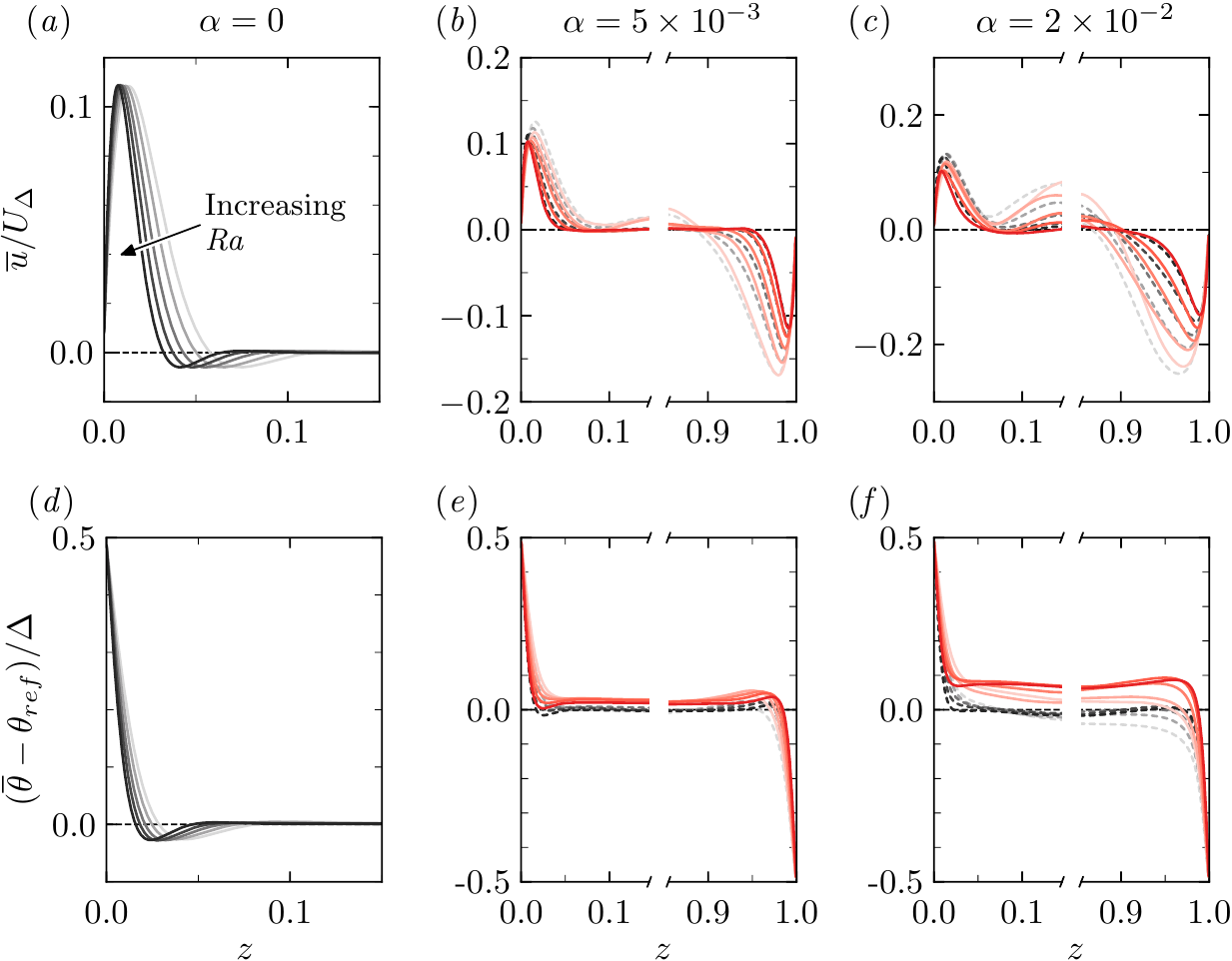}}
	\caption{\label{fig:MeanStatsInZ}Mean profiles as a function of horizontal location $z$: ($a$-$c$) vertical velocity, ($d$-$f$) temperature. ($a$,$d$) $\alpha=0$, ($b$,$e$) $\alpha=5\times 10^{-3}$, and ($c$,$f$) $\alpha=2\times 10^{-2}$. For $\alpha=0$, only the left half of the profiles are shown since the profiles are antisymmetric about the vertical centreline. Dashed grey curves represent mechanical coupling only. Solid red curves represent mechanical and thermal coupling. Darker curves represent higher $\Ray$.}
\end{figure}
Now, we turn our focus to the flow statistics. To establish a baseline, we first analyse the influence of the droplets on the mean flow \dl{profiles} of VC. 

\begin{figure}
	\centering
	\centerline{\includegraphics[trim=0 -0.5pc 0 -1pc,clip=true,scale=1]{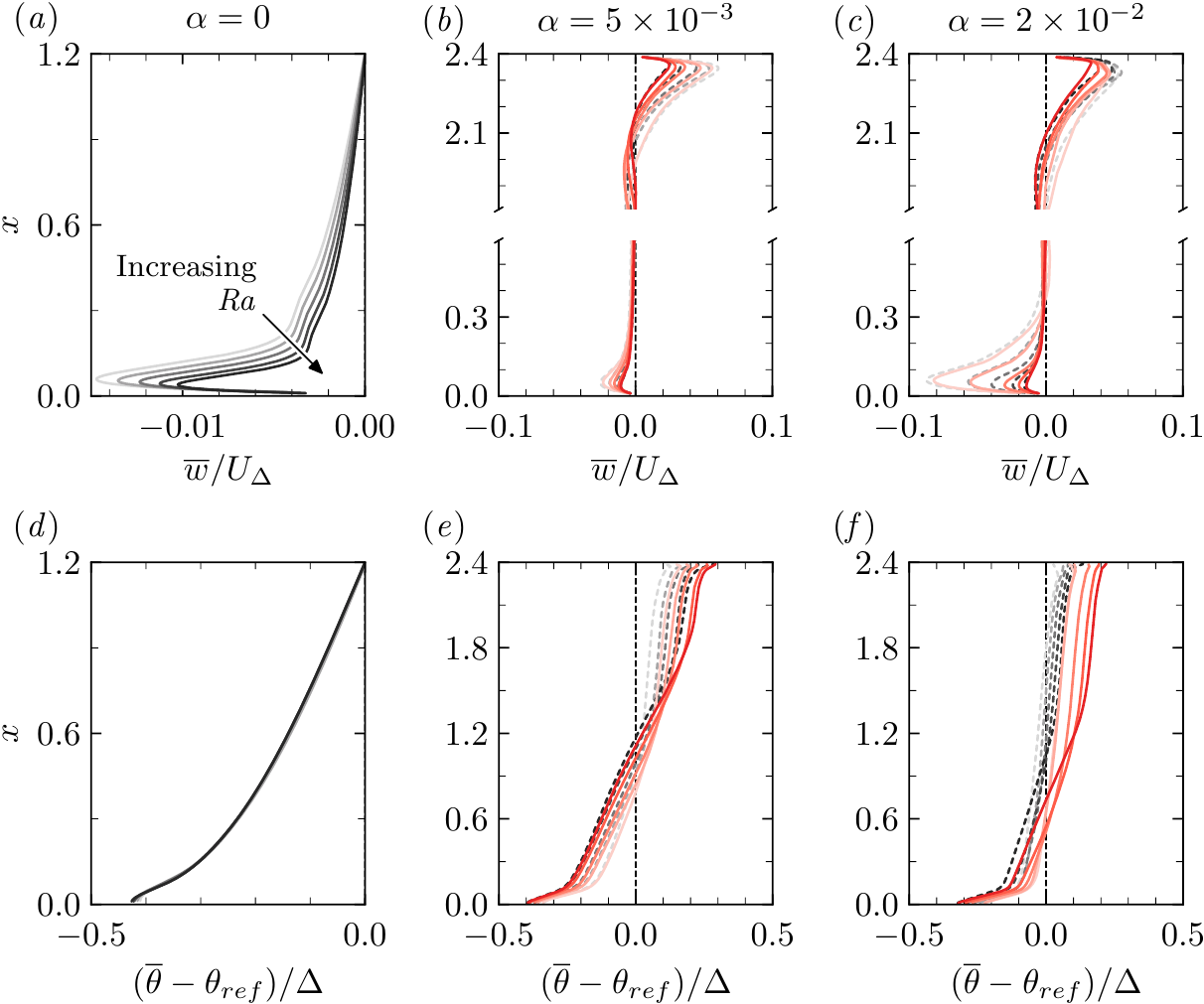}}
	\caption{\label{fig:MeanStatsInX}Same as figure \ref{fig:MeanStatsInZ}, but now for the mean profiles as a function of the vertical spatial variable $x$: ($a$-$c$) $\overline{w}/U_\Delta$, ($d$-$f$) $(\overline{\theta}-\theta_{\text{ref}})/\Delta$. For $\alpha = 0$, only the velocity profiles between $0\leqslant x \leqslant 1.2$ are shown since the profiles are antisymmetric about the horizontal centreline. Colour legends are the same as figure \ref{fig:MeanStatsInZ}.}
\end{figure}

Figure \ref{fig:MeanStatsInZ} shows the mean vertical velocity and temperature \dl{profiles} plotted versus $z$ (Note that all length scales have been made dimensionless with $L_z$). Without droplets, the mean \dl{profiles} are anti-symmetric about the channel centre-line (figure \ref{fig:MeanStatsInZ}$a$,$d$). The cell centre is stably stratified (figure \ref{fig:MeanStatsInX}$d$) with $\romd \overline{\theta}/\romd z|_{z=0.5} = 0$ and $\overline{u}|_{z=0.5}=0$. Therefore, unlike the doubly-periodic VC setup \citep{Ng+Ooi+Lohse+Chung.2014,Ng+Ooi+Lohse+Chung.2017}, there is no persistent mean shear in the bulk of the flow. For $\alpha > 0$ and for both coupling cases, the mean vertical velocity profiles are asymmetric with a much stronger downward velocity magnitudes near the cooler walls (figures \ref{fig:MeanStatsInZ}$b$,$c$). The difference between the maxima and minima of $\overline{u}$ is largest for the smallest $\Ray$, indicating that the droplet forcing is strongest.

The mean temperature profiles (figures \ref{fig:MeanStatsInZ}$e$,$f$) also exhibit asymmetries. For $\alpha = 5\times 10^{-3}$ and at the lowest $\Ray$, the temperature profiles for both coupling cases are relatively constant and do not exhibit any undershoot, which is observed for $\alpha = 0$ in figure \ref{fig:MeanStatsInZ}($d$) at $z\approx 0.04$. However, at higher $\Ray$, the profiles now bear some resemblance to the cases when $\alpha = 0$, corroborating the notion that thermal driving increasingly dominates. Here, we note that although $\overline{\theta} > \theta_{\text{ref}}$ in the bulk, the globally averaged temperature field $\langle \theta \rangle_t$ is statistically stationary within 0.5\% for all cases. In the bulk region ($0.2 \leqslant z \leqslant 0.8$), we obtain $\romd \overline{\theta}/\romd z|_{\text{bulk}}\approx 0$. Based on these results, the influence of the light droplets is seemingly most pronounced at the vertical boundaries as compared to the bulk.

\begin{figure}
	\centering
	\centerline{\includegraphics[trim=0 -0.5pc 0 -1pc,clip=true,scale=1]{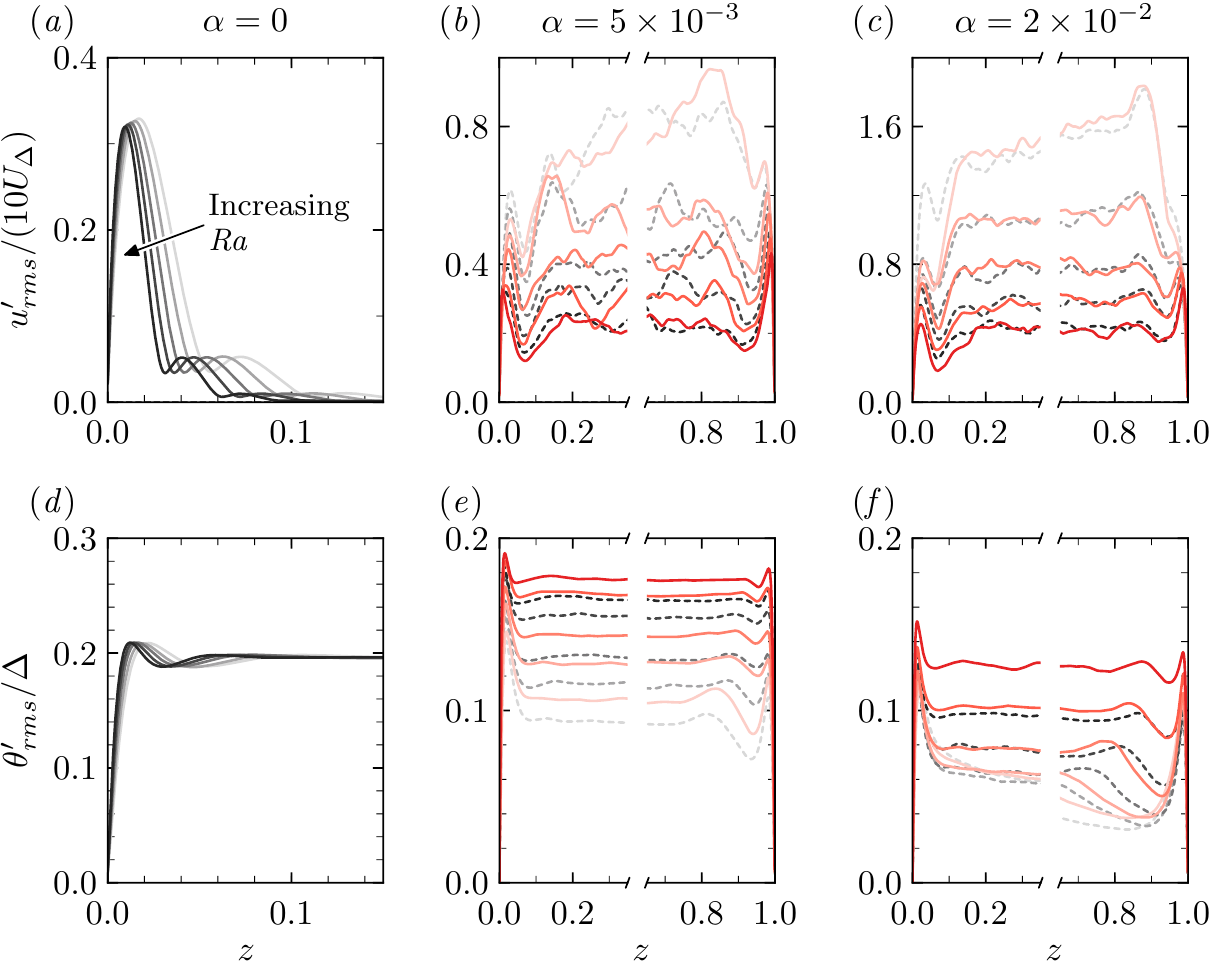}}
	\caption{\label{fig:TurbStatsInZ}Same as figure \ref{fig:MeanStatsInZ}, but now for the r.m.s. statistics as a function of $z$: ($a$-$c$) $u'_{rms}/(10U_\Delta)$, ($d$-$f$) $\theta'_{rms}/\Delta$. For $\alpha = 0$, only the left half of the profiles are shown since they are antisymmetric about the vertical centreline. Colour legends are the same as figure \ref{fig:MeanStatsInZ}.}
\end{figure}

\subsection{Profiles of mean horizontal velocity and temperature}
Figure \ref{fig:MeanStatsInX} shows the mean horizontal velocity and temperature \dl{profiles} plotted versus $x$. When $\alpha = 0$, the velocity profiles are antisymmetric (figure \ref{fig:MeanStatsInX}$a$) and the temperature profiles are constant for all $\Ray$ (figure \ref{fig:MeanStatsInX}$d$). When $\alpha > 0$, the antisymmetries are destroyed: for $\alpha = 0.5\times 10^{-2}$, the horizontal velocities are larger at the top wall (figure \ref{fig:MeanStatsInX}$b$), whereas for $\alpha = 2\times 10^{-2}$, the horizontal velocities are larger at the bottom wall (figure \ref{fig:MeanStatsInX}$c$). The source for the asymmetry can be traced to passage of droplets entering the bottom or leaving the top of the domain: \cs{at the lower boundary, the droplets which have near zero velocity block the horizontal flow causing the fluid to accelerate around the droplets. At the upper boundary, the droplets exit the domain at terminal velocity, and the entrained fluid impinges on the upper wall. Both mechanisms trigger intermittent intrusions of hotter and colder fluid at the upstream corners of the thermal boundary layers at the vertical walls. Since the blockage factor is higher for the $\alpha = 2\times 10^{-2}$ cases, the magnitude of the mean horizontal velocities are larger at $x \lesssim 0.3$ as compared to the $\alpha = 5\times 10^{-3}$ cases.}

For the temperature profiles, we note an overall weakening of the stable stratification at higher $\alpha$ (figures \ref{fig:MeanStatsInX}$e$ and $f$), with the bulk mean temperatures $\overline{\theta} \rightarrow \theta_{\text{ref}}$. The relatively uniform value of $\overline{\theta}$ for the most part of $x$ indicates strong mixing of the thermal field with increasing $\alpha$. 

\subsection{r.m.s. profiles of vertical velocity and temperature} \label{subsec:RMSProfiles}

The root-mean-square (r.m.s.) of the fluctuating quantities are plotted in figure \ref{fig:TurbStatsInZ} for all cases as a function of horizontal distance $z$. Here, we define $(\cdot)'_{rms}\equiv [(\overline{u'})^2]^{1/2}$. When $\alpha = 0$ (figure \ref{fig:TurbStatsInZ}$a$,$d$), both $u'_{rms}$ and $\theta'_{rms}$ exhibit near wall peaks and are symmetrical about the channel-centreline.

When $\alpha > 0$, the bulk velocity fluctuations $u'_{\text{bulk},rms} > 0$ as a direct result of droplet induced liquid fluctuations. Interestingly, $u'_{\text{bulk},rms}$ at lower $\Ray$ values are much larger than at higher $\Ray$, which highlights the larger influence of droplet forcing on the flow at lower $\Ray$. The $u'_{rms}$ profiles also exhibit slight asymmetry with values tending to be larger closer to the colder wall as compared to the hotter wall. This asymmetry is consistent with the notion of a more intermittent colder downwards flow caused by the disruption of the large-scale circulation by the droplet passage, as discussed in \S\,\ref{subsec:MeanProfiles}.
 
For $\theta'_{\text{rms}}$, the magnitudes in the bulk for $\alpha > 0$ (figures \ref{fig:TurbStatsInZ}$e$,$f$) tend to be lower than for the case when $\alpha = 0$ (figure \ref{fig:TurbStatsInZ}$d$), where $\theta'_{\text{rms},\text{bulk}} \approx 0.2$. With thermal coupling, the $\theta'_{\text{rms}}$ profiles are typically slightly larger than without thermal coupling and counteracts the mechanical agitation by the droplets. This effect can be explained by the thermal exchange of the droplet and the surrounding liquid which induces local thermal fluctuations. Therefore, both the mechanical agitation at larger $\alpha$ and the thermal coupling of the droplets contribute to the bulk mixing of the thermal field.

\subsection{Scaling of Nusselt and Reynolds numbers versus Rayleigh number} \label{subsec:NuReRa}
\begin{figure}
	\centering
	\centerline{\includegraphics[trim=0 -0.5pc 0 -1pc,clip=true,scale=1]{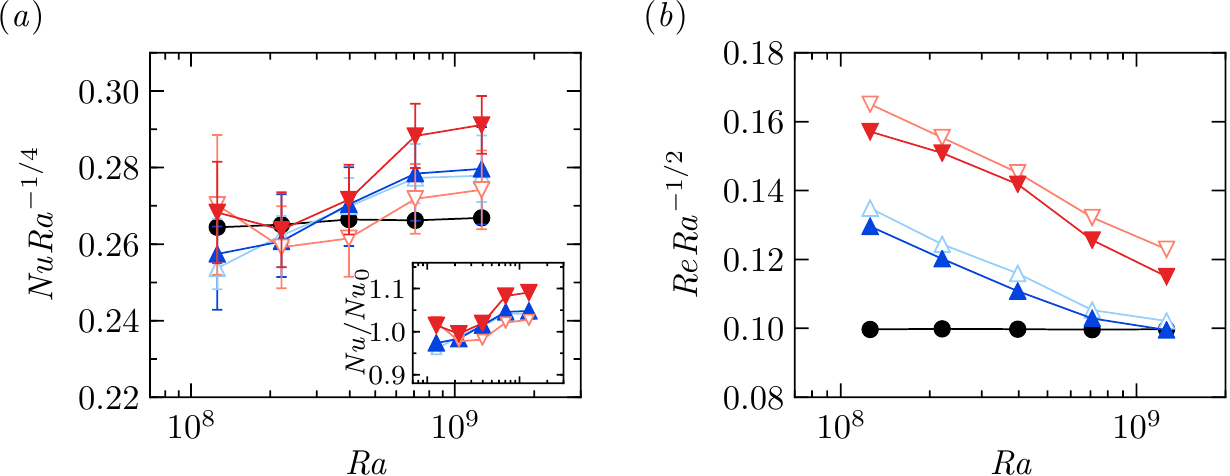}}
	\caption{\label{fig:NuRaComp}($a$) Compensated $\Nus$ versus $\Ray$, and ($b$) compensated $\Rey$ versus $\Ray$. Inset of ($a$): ratio of $\Nus|_{\alpha > 0}$ to $\Nus|_{\alpha = 0}$. Solid black symbols are DNS data for $\alpha = 0$. Upwards-pointing blue triangles are for $\alpha = 5\times 10^{-3}$ cases and downwards-pointing red triangles are for $\alpha = 2\times 10^{-2}$. Open triangles denote mechanical coupling only and filled triangles denote both mechanical and thermal coupling. The $\Nus$ versus $\Ray$ and $\Rey$ versus $\Ray$ scalings for $\alpha = 0$ are consistent with analytical predictions for VC driven by laminar boundary layers, \ie\,$\Nus\sim\Ray^{1/4}$ and $\Rey\sim\Ray^{1/2}$ at constant $\Pran$ \citep{Shishkina.2016.momentum}.}
\end{figure}
In figure \ref{fig:NuRaComp}, we present the scaling of the $\Nus$ and $\Rey$ versus $\Ray$. Here, we employ the wind-based Reynolds number, $\Rey \equiv \overline{u}_{\text{max}}L_z/\nu$ as a measure of the large scale circulation.

When $\alpha = 0.0$ (solid circles, figure \ref{fig:NuRaComp}), we find that $\Nus \sim \Ray^{0.25\pm0.003}$ and $\Rey \sim \Ray^{0.50\pm0.002}$ which are in agreement with the $\Nus \sim \Ray^{1/4}$ and $\Rey \sim \Ray^{1/2}$ analytical predictions for laminar boundary layer-dominated VC \citep{Shishkina.2016.momentum}. For pure mechanical coupling (open triangles), the $\Nus$ trends exhibit steeper slopes and from a least-square fit to a power-law, we obtain $\Nus \sim \Ray^{0.29\pm0.02}$ for $\alpha = 5\times 10^{-3}$ and $\Nus \sim \Ray^{0.26\pm0.04}$ for $\alpha = 2\times10^{-2}$. In contrast, the $\Rey$ slopes are less steep and from a least-square fit to a power-law, we obtain $\Rey \sim \Ray^{0.38\pm0.03}$ for $\alpha = 5\times 10^{-3}$ and $\Rey \sim \Ray^{0.37\pm0.02}$ for $\alpha = 2\times10^{-2}$. When both mechanical and thermal coupling are enabled (filled triangles), we obtain $\Nus \sim \Ray^{0.29\pm0.02}$ and $\Rey \sim \Ray^{0.38\pm0.03}$ for $\alpha = 5\times 10^{-3}$, and $\Nus \sim \Ray^{0.29\pm0.04}$ and $\Rey \sim \Ray^{0.36\pm0.06}$ for $\alpha = 2\times10^{-2}$. When comparing the coupling cases, the effective scaling for $\Nus$ and $\Rey$ is largely unaffected. However, by including thermal coupling, the temperature field is distributed more efficiently and so, the magnitude of the heat transport is increased. 

As a direct comparison for $\Nus$, the ratio $\Nus/\Nus_{\alpha=0}$ is shown in the inset of figure \ref{fig:NuRaComp}($a$) \cs{and the values range \dl{from} 0.95 to 1.1. Some caution is warranted here when interpreting the ratios. Because of the rather large variations of $\Nus\Ray^{-1/4}$ as shown in the figure, we cannot conclusively claim that there exist a decrease in $\Nus$ at low $\Ray$. However, we can link the variations of the ratios to the different manner in which the droplets locally influence the wall heat fluxes and wall shear stresses. The local influences are quantified and discussed in \S\,\ref{sec:LocalNu} and in \S\,\ref{sec:LocalCf}.}

Now, we focus on the $\Rey$ trends. For $\alpha > 0$, the $\Rey$ values tend to be larger than \dl{for} the $\alpha = 0$ case and this is consistent with the response of the VC flow due to the passage of the droplets across the top and bottom boundary layers. As the droplets cross the \cs{horizontal} boundary layers, the large-scale circulation of the background VC flow is continuously disrupted, triggering horizontal intrusions of warmer fluid at the top wall and cooler fluid at the bottom wall (peaks in mean horizontal velocities in figures \ref{fig:MeanStatsInX}$b$ and $c$), similar to the intrusions observed in transient VC in a square cavity \citep{Patterson+Imberger.1980,Armfield+Patterson.1991}. For $\alpha = 5\times 10^{-3}$, at the higher $\Ray$-values, the $\Rey$-values tend to approach the $\Rey$ values for $\alpha = 0$. This incipient trend suggests that the droplet driving is no longer dominant at this \dl{part of the} parameter space as compared to the $\alpha = 2\times 10^{-2}$ case.

\begin{figure}
	\centering
	\centerline{\includegraphics[trim=0 -0.5pc 0 -1pc,clip=true,scale=1]{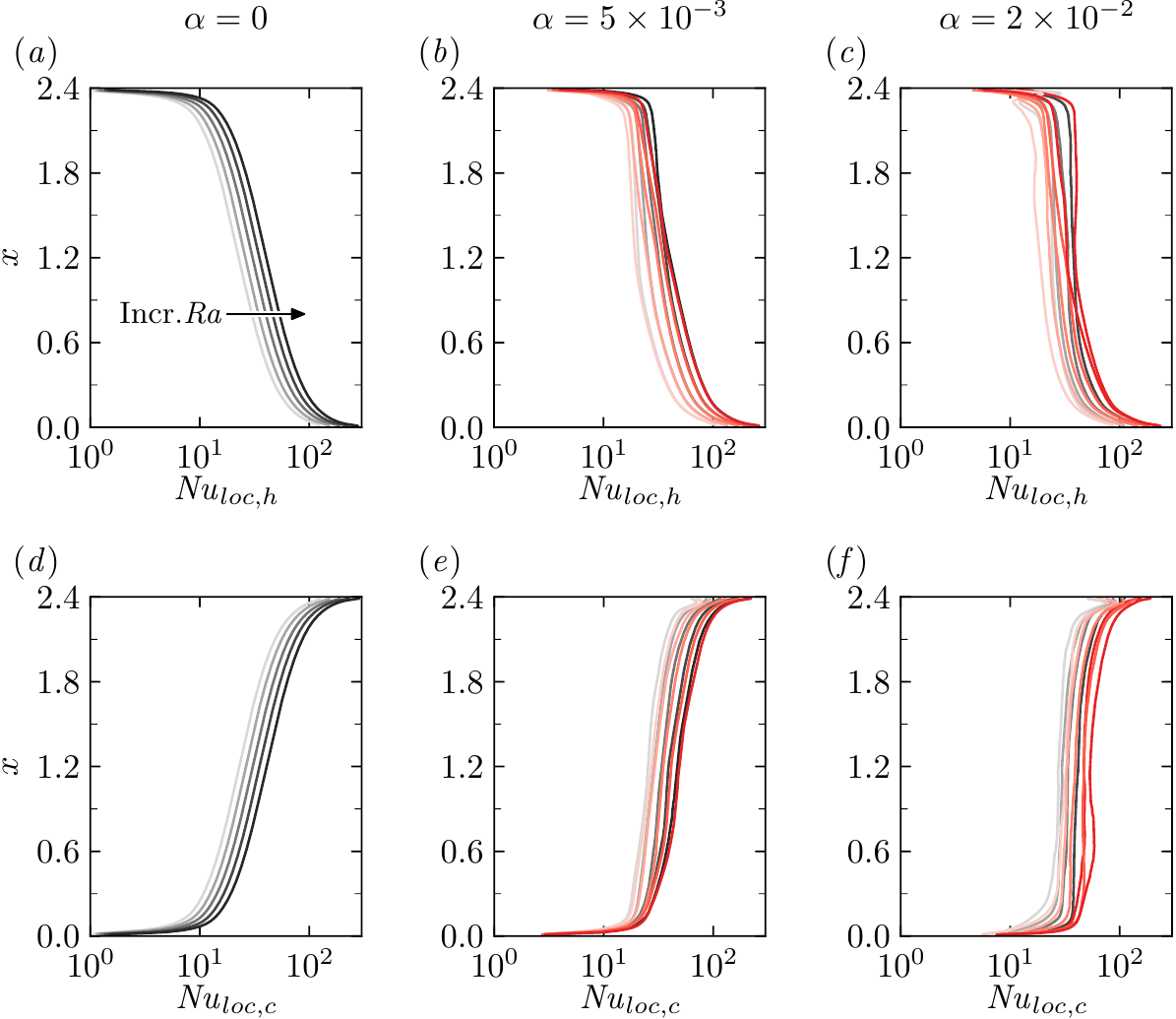}}
	\caption{\label{fig:LocalNuTrends}Profiles of $\Nus_{\text{loc}}$ plotted as a function of $x$ at the hot wall ($a$-$c$), and the cold wall ($d$-$f$). Colour legends are the same as figure \ref{fig:MeanStatsInZ}.}
\end{figure}

\section{Droplet influence on local Nusselt number} \label{sec:LocalNu}

In this section, we link the $\Nus$ versus $\Ray$ variations discussed in \S\,\ref{subsec:NuReRa} to the changes in the local Nusselt number evaluated at the hot and cold walls. We define the local Nusselt number as $\Nus_{\text{loc}} \equiv f_{w,\text{loc}}L_z/(\Delta \kappa) = [\partial \overline{\theta}(x)/\partial z]|_w/(\Delta L_z)$, which is the local dimensionless temperature gradient evaluated at $z=0$ and $L_z$. The trends are shown in figure \ref{fig:LocalNuTrends} as function of $x$.

From figure \ref{fig:LocalNuTrends}, $\Nus_{\text{loc}}$ are larger in the upstream of the vertical boundary layers, that is $x \lesssim 1.2$ for figures \ref{fig:LocalNuTrends}($a$-$c$) and $x \gtrsim 1.2$ for figures \ref{fig:LocalNuTrends}($d$-$f$). Here, the larger values of $\Nus_{\text{loc}}$ simply reflect the thinner thermal boundary layers developing from the corners of the domain. For $\alpha = 0$, $\Nus_{\text{loc}}$ monotonically decreases as the boundary layer develops and is consistent across the $\Ray$ range. However, the trends vary considerably for $\alpha > 0$. For example, relative to the $\alpha = 0$ cases, (i) $\Nus_{\text{loc},h}$ becomes lower for $x \lesssim 1.2$, and (ii) for $\alpha = 2\times 10^{-2}$, both $\Nus_{\text{loc},h}$ and $\Nus_{\text{loc},c}$ are roughly constant for $0.6 \lesssim x \lesssim 1.8$. Since these changes directly reflect the thermal boundary layer thicknesses, we can conclude that the droplets not only influence the bulk statistics as shown in \S\,\ref{sec:SimulationResults}, but would also influence the local thermal boundary layers. 

\begin{figure}
	\centering
	\centerline{\includegraphics[trim=0 -0.5pc 0 -1pc,clip=true,scale=0.9]{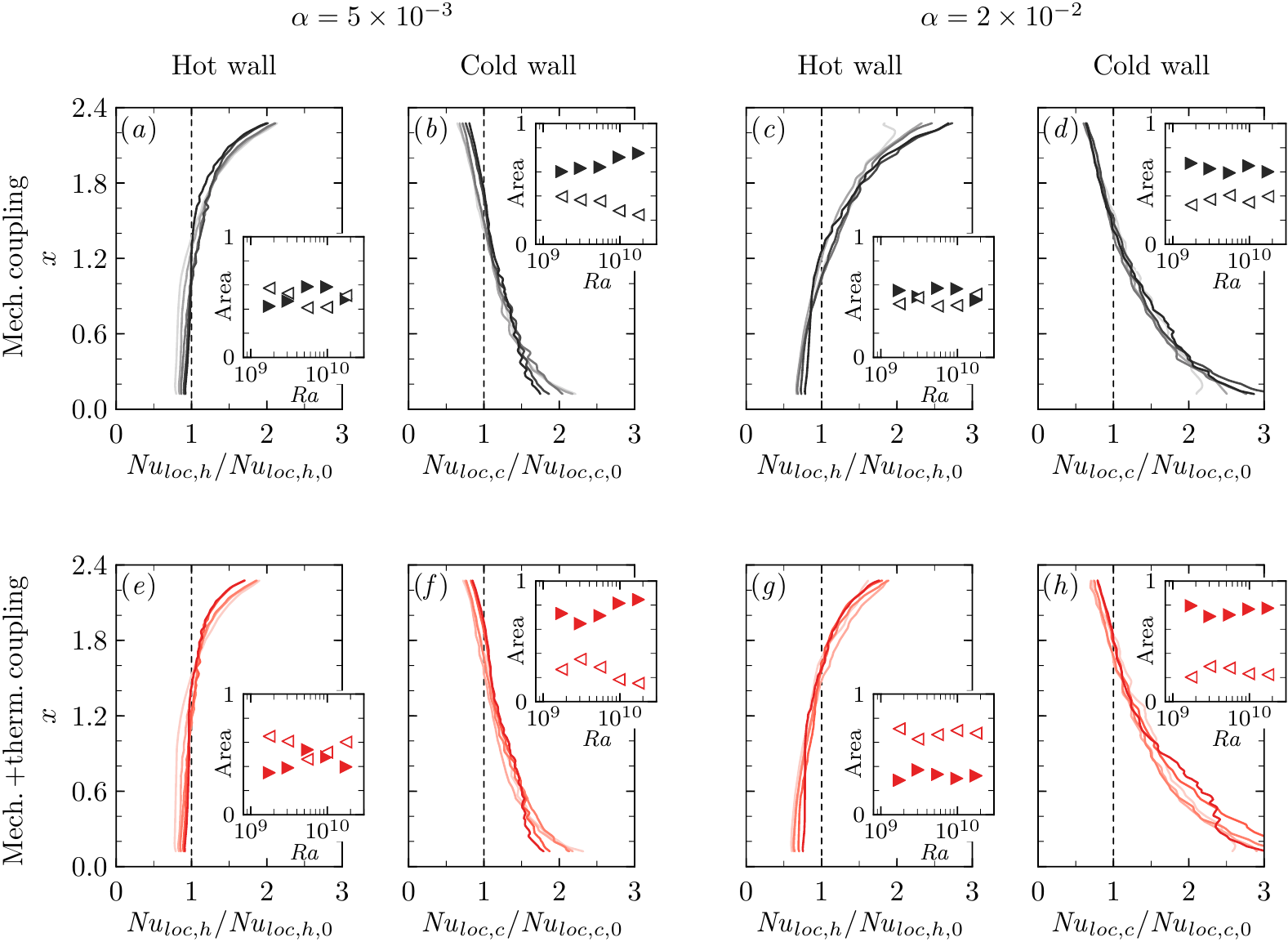}}
	\caption{\label{fig:LocalNuRatio}Ratio of $\Nus_{\text{loc}}$ to $\Nus_{\text{loc},0}$, which is the case for $\alpha = 0$. The corresponding wall-areas of the ratios versus $\Ray$ are shown in the inset: wall-areas with reduced $\Nus_{\text{loc}}$ are denoted by left-pointing open triangles, and increased $\Nus_{\text{loc}}$ by right-pointing solid triangles. Darker curves represent higher $\Ray$.}
\end{figure}
To emphasise the changes in $\Nus_{\text{loc}}$, we plot the ratio of $\Nus_{\text{loc}}$ and $\Nus_{\text{loc},0}$ in figure \ref{fig:LocalNuRatio}. ($\Nus_{\text{loc},0}$ is $\Nus_{\text{loc}}$ computed for the $\alpha = 0$ cases). The corresponding wall-areas for $\Nus_{\text{loc}}$ are also shown in the insets, with reduced-$\Nus_{\text{loc}}$ values denoted by left-pointing open triangles, and increased-$\Nus_{\text{loc}}$ values denoted by right-pointing solid triangles. For $\alpha = 5\times 10^{-3}$, the decreased $\Nus_{\text{loc},h}$ can be clearly seen for all $\Ray$ and $x/L_x \lesssim 1.2$ (figures \ref{fig:LocalNuRatio}$a$,$e$). This decreasing behaviour can also be observed for $\alpha = 2\times10^{-2}$, although the corresponding wall-area with decreased $\Nus_{\text{loc},h}$ is smaller for the mechanically coupled case (see figure \ref{fig:LocalNuRatio}$c$ and the inset plot). The decreased $\Nus_{\text{loc},h}$ for the $\alpha = 5\times 10^{-3}$ case overwhelms the increased $\Nus_{\text{loc},c}$ for $x/L_x \lesssim 1.2$, with the lowest $\Ray$ cases being most strongly influenced, as previously shown in figure \ref{fig:NuRaComp}. In contrast, $\Nus_{\text{loc},c}$ is significantly increased for $\alpha = 2\times 10^{-2}$ and $x \lesssim 1.2$ by roughly a factor of 1.5 times (figures \ref{fig:LocalNuRatio}$d$,$h$). Based on the much stronger droplet driving for $\alpha = 2\times 10^{-2}$, $\Nus|_{\alpha = 2\times 10^{-2}}$ is increased by about 5\% for the lowest $\Ray$ relative to $\Nus|_{\alpha = 5\times 10^{-3}}$. 

For the different distributions of $\Nus_{\text{loc}}$ in figures \ref{fig:LocalNuTrends} and \ref{fig:LocalNuRatio}, we emphasize \dl{that} they are non-universal phenomena for the flow problem considered since the mechanism of the changes rely on the strength of the droplet driving, $\Ray_d$. These changes therefore cannot be trivially determined \textit{a priori}. What can be discerned from the current results is that the droplets influence the bulk flow (as seen in the mean and r.m.s.\,statistics in figures \ref{fig:MeanStatsInZ} to \ref{fig:TurbStatsInZ}), the near-wall flow and the large-scale circulation of VC. Different mechanisms in these regions compete and the prevailing mechanism(s) would presumably determine the heat transport of the setup. 

\section{Droplet influence on local skin-friction coefficient}\label{sec:LocalCf}
\begin{figure}
	\centering
	\centerline{\includegraphics[trim=0 -0.5pc 0 -1pc,clip=true,scale=1]{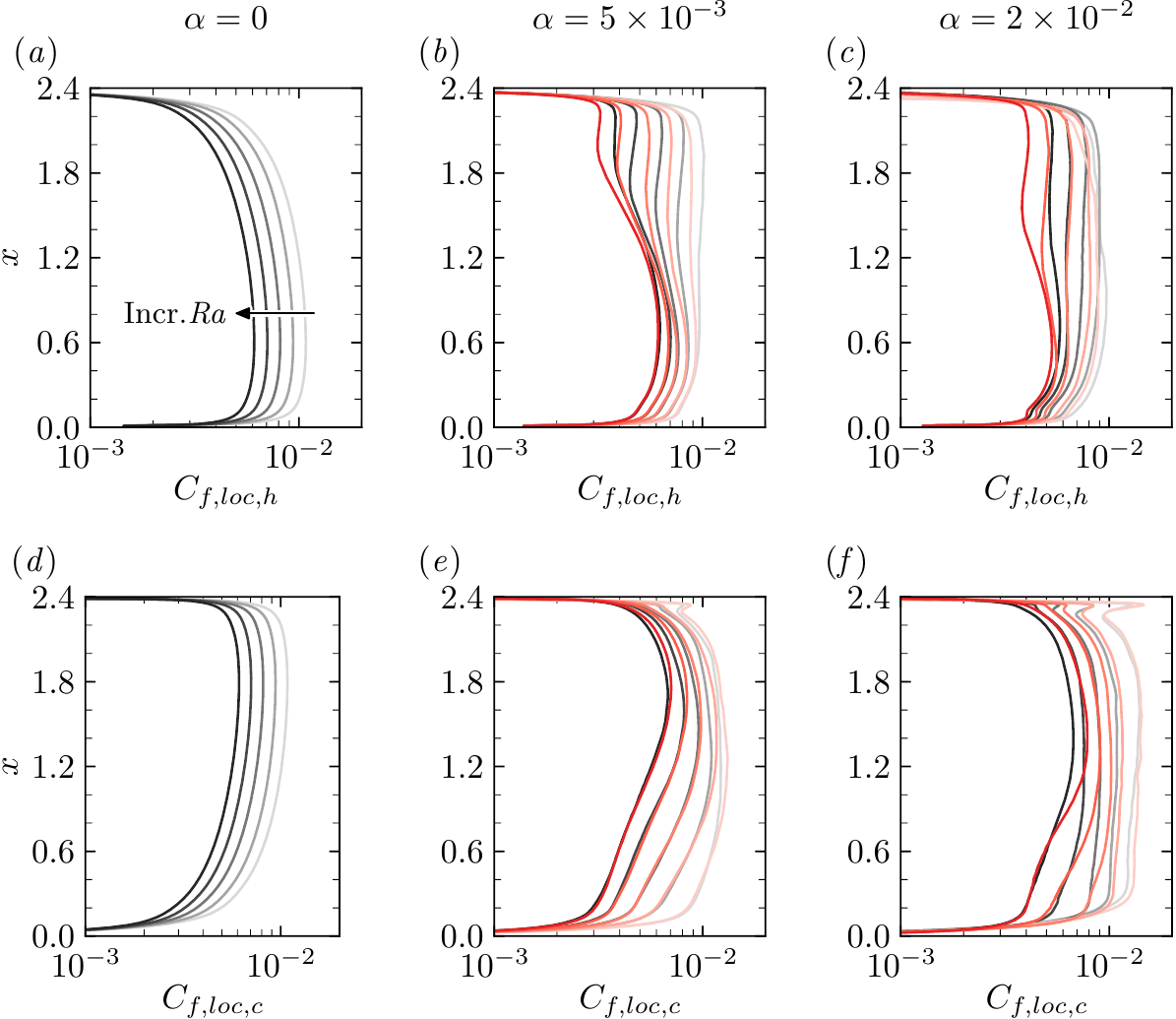}}
	\caption{\label{fig:LocalCfTrends}Similar to figure \ref{fig:LocalNuTrends}, but now for the local skin-friction coefficient $C_{f,\text{loc}}$. Colour legends are the same as figure \ref{fig:MeanStatsInZ}.}
\end{figure}

\begin{figure}
	\centering
	\centerline{\includegraphics[trim=0 -0.5pc 0 -1pc,clip=true,scale=0.9]{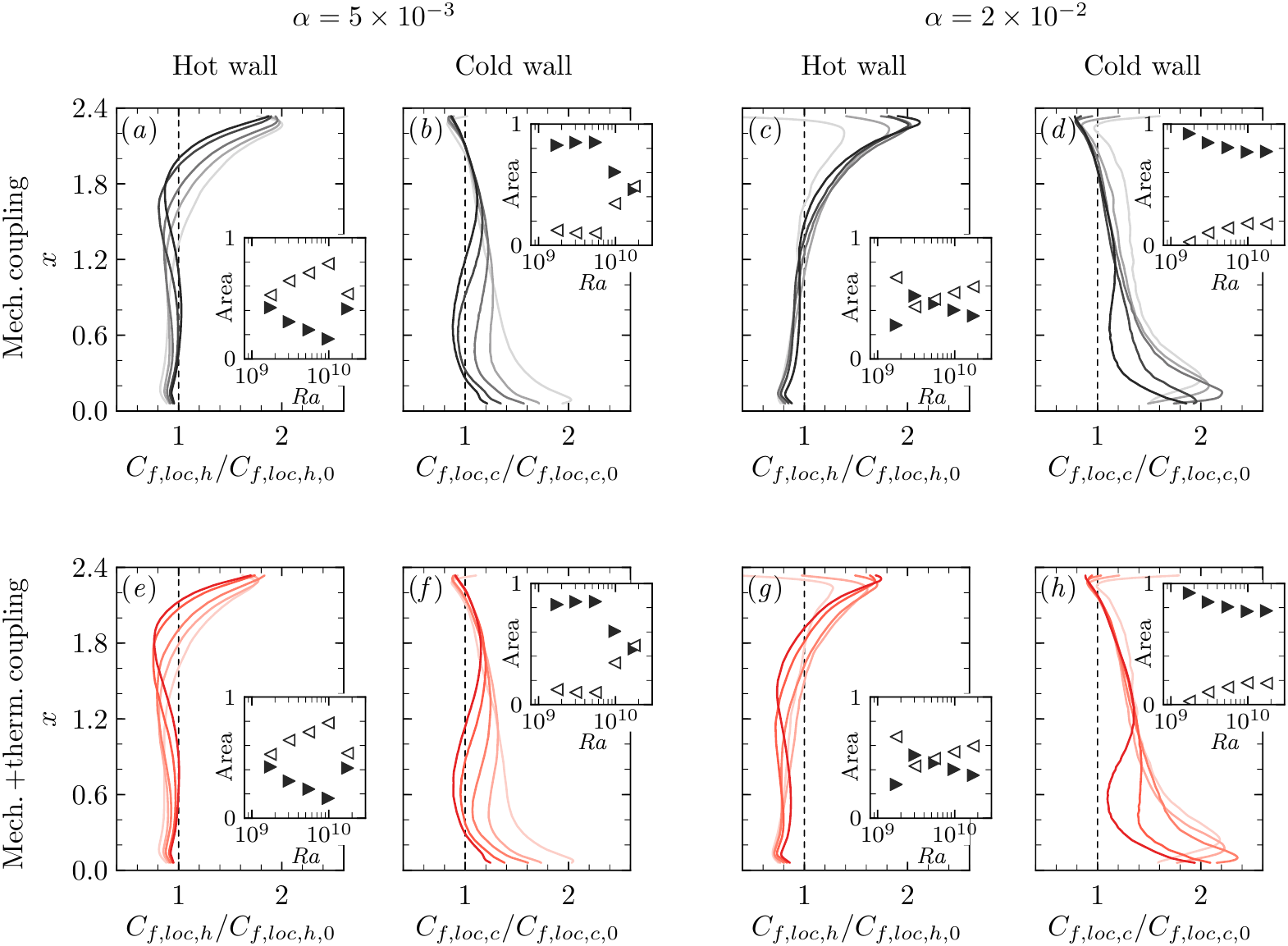}}
	\caption{\label{fig:LocalCfRatio}Similar to figure \ref{fig:LocalNuRatio}, but now for the local skin-friction coefficient $C_{f,\text{loc}}$. The corresponding wall-areas of the ratios versus $\Ray$ are shown in the inset: wall-areas with reduced $C_{f,\text{loc}}$ are denoted by left-pointing open triangles, and increased $C_{f,\text{loc}}$ by right-pointing solid triangles. Colour legends are the same as figure \ref{fig:LocalNuRatio}.}
\end{figure}

Unlike Rayleigh--Ben{\'a}rd convection, the thermal boundary layers in VC \cs{are sheared by a mean wind with a constant direction that is predetermined by the boundaries} \citep{Ng+Ooi+Lohse+Chung.2014}. Therefore, to quantify the influence of the droplets on wind-shearing, we plot the local skin-friction coefficient $C_{f,\text{loc}}$ versus $x$ in figure \ref{fig:LocalCfTrends}. Here, $C_{f,\text{loc}} \equiv 2\tau_{w}(x)/U_\Delta^2$, where $\tau_w(x) \equiv \sqrt{\mu \partial \overline{u}(x)/\partial z|_w}$ is the wall shear stress. Similar to the idea of figure \ref{fig:LocalNuRatio}, \cs{the relative changes in the local skin-friction coefficients are plotted in figure \ref{fig:LocalCfRatio}.}

For $\alpha = 0$, $C_{f,\text{loc}}$ is largest at wall heights that are close to the upstream of the developing boundary layer. However, when $\alpha > 0$, $C_{f,\text{loc}}$ \dl{is} roughly constant for the most part of $x$ at low $\Ray$. Two points can be made from the distributions of $C_{f,\text{loc}}$. First, the roughly uniform distribution of $C_{f,\text{loc}}$ at low $\Ray$ for $\alpha>0$ imply that the droplet driving dominates the mean wind of VC and, on a mean sense, homogenizes the viscous boundary layer particularly at the hot wall. Second, the distributions of $C_{f,\text{loc}}$ are not symmetric at the hot and cold walls (for example, max[$C_{f,\text{loc},c}$]~$>$~max[$C_{f,\text{loc},h}$]) as compared to the $\alpha=0$ case (figure \ref{fig:LocalNuRatio}$a$ and $d$). One possible explanation of this asymmetry can be made by observing the rising direction of the droplets: At the cold wall, the droplets oppose the downwards flow whereas at the hot wall, the droplets aid the upwards flow. Coupled with the asymmetry of the mean horizontal velocity profiles in figure \ref{fig:MeanStatsInX}, the resulting viscous boundary layer becomes thinner at the cold wall, and a larger $C_{f,\text{loc}}$ \dl{results}. However, this conjecture may not hold at higher $\Ray$ cases because the viscous boundary layers eventually become much thinner and closer to the walls. As a result, at sufficiently high $\Ray$, the influence of droplets presumably diminishes with increasing distance from the edge of the viscous boundary layers, eventually yielding to the dynamics of thermal driving.

When compared with $C_{f,\text{loc},0}$ (figure \ref{fig:LocalCfRatio}), we find larger values of $C_{f,\text{loc}}$ in concomitant regions with larger values of $\Nus_{\text{loc}}$ in figure \ref{fig:LocalNuRatio}. Interestingly, whilst $\Nus_{\text{loc}}$ is relatively insensitive to $\Ray$ (see figure \ref{fig:LocalNuRatio}), the wall-height distributions of $C_{f,\text{loc}}$ exhibit a strong non-monotonic behaviour which depends \dl{on} $\Ray$, $\alpha$ and whether the cold or hot wall is considered. Therefore, it appears that $C_{f,\text{loc}}$ is more sensitive to the droplets induced agitation as compared to $\Nus_{\text{loc}}$. These results provide a strong indication that the light droplets interact with VC flow in a non-universal manner.

\section{Light droplets versus bubbles - a comparison to experiments by \cite{Gvozdic+Others.2018}} \label{sec:ComparisonWithExp}

\begin{figure}
	\centerline{\includegraphics[trim=0 -0.5pc 0 -1pc,clip=true,scale=1]{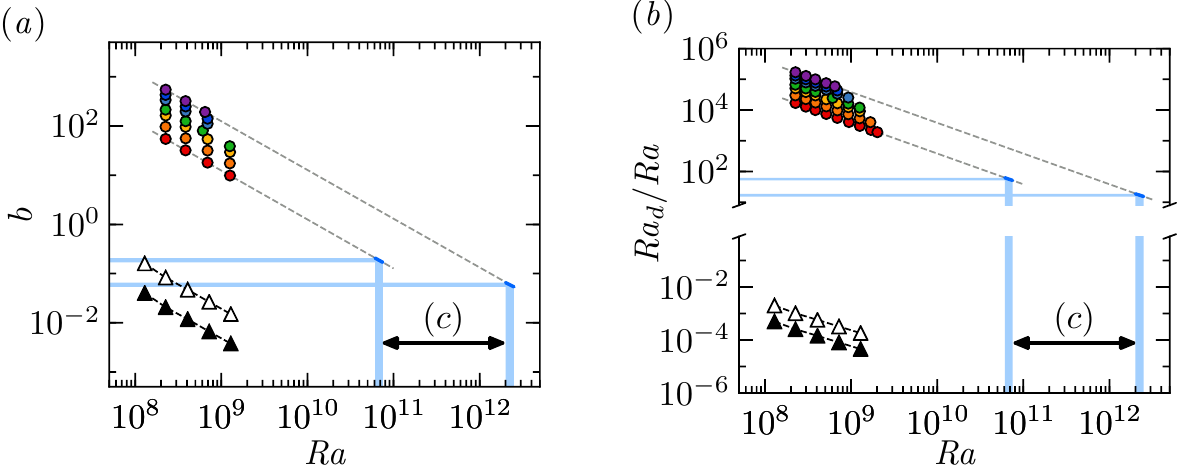}}
	\caption{\label{fig:RaCritPrediction}Plots of ($a$) bubblance parameter, $b$, versus $\Ray$, and ($b$) $\Ray_d/\Ray$ versus $\Ray$. The range of $\Ray_c(\alpha)$ values from the experiments of \cite{Gvozdic+Others.2018} are marked in ($c$). Present DNS results are solid triangles for $\alpha=5\times10^{-3}$ and open triangles for $2\times 10^{-2}$. Experimental values (circles, red to purple fill denoting increasing $\alpha$) are estimated from the physical properties reported in \cite{Gvozdic+Others.2018}. In ($a$), only a subset of the experimental values are estimated by assuming values of $u_0$ at matched $\Ray$ to our DNS results. The intersect of ($c$) and the dashed grey lines, \ie\,$b\sim\Ray_c^{-1}$ and $\Ray_d/\Ray\sim\Ray_c^{-1}$, are approximations of $b$ and $\Ray_d/\Ray$ for the lowest and highest $\alpha$ from laboratory experiments.}
\end{figure}
In this section, we discuss several aspects of the physical parameters in our simulations, which distinguish our findings from the laboratory results of bubbly VC by \cite{Gvozdic+Others.2018}. 

A crucial difference between our investigation and the experiments is that $\hat{\rho} = 0.5$ in our simulations (corresponding to light droplets) whereas $\hat{\rho}\approx 10^{-3}$ in their experiments (corresponding to air bubbles in water). Clearly, the large differences in the density ratios play a role and this is reflected in our simulations. For example, the mean temperature in the bulk region of our simulations have approximately zero gradient (figure \ref{fig:MeanStatsInZ}), whereas the mean temperature in the bulk region of the experiments have a finite gradient (see figure 9($a$) of their paper), indicating a much stronger mixing of the thermal field by the bubbles as compared to light droplets. Furthermore, the values of $\Nus$ for VC with light droplets is within 10\% of the $\Nus$ values without droplets (figure \ref{fig:NuRaComp}), whereas in the laboratory experiments of \cite{Gvozdic+Others.2018}, $\Nus$ can be larger by up to 20 times with bubbles than without and remain $\Ray$-independent for their investigated parameter range. Therefore, we conclude that the background VC flow remains relatively dominant even with influence of light droplets, and this is reflected in the non-universal distributions of the local heat transport and skin-friction coefficients shown in figures \ref{fig:LocalNuRatio} and \ref{fig:LocalCfRatio}. 

The strength of the bubble-induced agitation versus droplet-inducted agitation can also be quantified \textit{a posteriori} using the bubblance parameter,
\begin{equation}
b \equiv U_b^2\alpha/u_0^2, \label{eqn:bubblanceparam}
\end{equation}
\citep[\cf][]{Lance+Bataille.1991,vanWijngaarden.1998,Rensen+Luther+Lohse.2005,Almeras+Mathai+Lohse+Sun.2017}, which defines the ratio of energy produced by a bubble swarm, \ie\,$U_b^2\alpha$, and the energy of the background turbulence without bubbles, \ie\,$u_0^2$. Note that a prefactor of 1 is chosen for (\ref{eqn:bubblanceparam}), which is different to previous definitions which employ a prefactor of $1/2$ \citep[based on the added mass coefficient, \cf][]{Rensen+Luther+Lohse.2005}, however the present discussions are still valid. We define $U_b$ as the mean bubble or droplet rise velocity and $u_0$ as the maximum of the mean vertical velocity of the single phase flow at the half-height of the domain. Next, we estimate $b$ for our DNS and for the experiments by \cite{Gvozdic+Others.2018}.

Since we have the full information from our DNS, the calculation of $b$ is straightforward. For the laboratory experiments, $u_0$ was not recorded and so, invoking dynamic similarity, we estimate the values using our DNS results at matched $\Ray$. $U_b$ is assumed to be $0.34$\,\mbox{m\,s$^{-1}$} for the laboratory experiments. The values of $b$ are plotted in figure \ref{fig:RaCritPrediction}($a$). From the figure, we find that $O(10^{-3})\lesssim b \lesssim O(10^{-1})$ for our DNS whereas $O(10^1) \lesssim b\lesssim O(10^3)$ for the experiments. The much smaller magnitude of $b$ for our DNS clearly indicates that light droplets produce much lower kinetic energy compared bubbles. Also, $b$ decreases with increasing $\Ray$ and implies that the kinetic energy of the background flow will eventually dominate the (constant) injection of kinetic energy by the dispersed phase. Based on the same idea, we compare the ratio of $\Ray_d/\Ray$ for our DNS and the experiments in figure \ref{fig:RaCritPrediction}($b$). From the figure, we observe a similar scale separation and decreasing trend with increasing $\Ray$: The values are $O(10^{-5}) \lesssim \Ray_d/\Ray \lesssim O(10^{-3})$ for our DNS and $O(10^{3}) \lesssim \Ray_d/\Ray \lesssim O(10^{5})$ for the experiments, confirming that the bubble driving is indeed a stronger driving mechanism than light droplets. 

It is useful for applications such as in chemical mixing to have an estimate of the parameter space for $b$ or $\Ray_d/\Ray$ where the driving by background turbulence eventually dominates bubble driving. For the laboratory experiments with bubbly VC, \cite{Gvozdic+Others.2018} estimated this parameter space by defining a critical Rayleigh number, $\Ray_c$, as follows:  First, an effective power-law trend of $\Nus \sim \Ray^{0.33}$ is obtained from the single phase experiments. Then, observing that the $\Nus$ trends are insensitive to $\Ray$ for $5\times10^{-3} \lesssim \alpha \lesssim 5\times10^{-2}$ \citep[\cf\,figure 12 of][]{Gvozdic+Others.2018}, the $\Nus \sim \Ray^{0.33}$ and constant $\Nus$ trends are extrapolated to higher $\Ray$ values. The intersection of these curves are defined as $\Ray_c$, where $7\times10^{10} \lesssim \Ray_c(\alpha) \lesssim 2\times10^{12}$ for the $\alpha$ values investigated. The range of $\Ray_c$ values are marked in figure \ref{fig:RaCritPrediction}($c$).

We can now directly extrapolate the trends of $b$ and $\Ray_d/\Ray$ to the $\Ray_c$ values. From least-square fits, the effective power laws are $b\sim\Ray^{-1}$ and $\Ray_d/\Ray \sim \Ray^{-1}$. Therefore, the extrapolated values are $b\sim\Ray_c^{-1}$ and $\Ray_d/\Ray \sim \Ray_c^{-1}$, visually marked by the blue patches in figure \ref{fig:RaCritPrediction}. For illustration purposes, only the $\alpha = 5\times10^{-3}$ and $5\times10^{-2}$ are drawn and an allowance of $\Ray_c\pm 10\%$ was employed to compute the extrapolation. The corresponding values are $(b,\Ray_d/\Ray)|_{\alpha=5\times 10^{-3}} \approx (0.2,60)$ and $(b,\Ray_d/\Ray)|_{\alpha=5\times 10^{-2}} \approx (0.06,18)$. These values suggest that the VC flow will dominate bubble-induced liquid agitation at $b \lesssim O(10^{-1})$ and $\Ray_d/\Ray \lesssim O(100)$. We note that our dataset for $\alpha=2\times10^{-2}$ coincide with this regime for $b|_{\alpha=5\times 10^{-2}}$ (lower horizontal blue line in figure \ref{fig:RaCritPrediction}$a$), however, since the boundary layer dynamics are still dominant for our configuration, it suggests that $\hat{\rho}$ is an additionally important parameter when characterising bubbly turbulence. Interestingly, for bubbles \dl{rising in grid-generated turbulence (or incident turbulence)}, \cite{Almeras+Mathai+Lohse+Sun.2017} determined a slightly larger value for $b$ ($\approx 0.7$), where bubble-induced agitation appears to dominate. The mechanism was related to an increase in development length of the secondary bubble wake, which significantly enhances liquid velocity fluctuations. Indeed, the values of $b$ from our DNS are smaller which is consistent with the notion that the background flow remains dominant for our parameter space considered.

\section{Conclusions and outlook} \label{sec:ConclusionsOutlook}

In this study, we simulated the VC flow with dispersed light droplets between $\Ray = 1.3\times 10^{8}$ and $1.3\times 10^{9}$ and $\Pran$-value of 7. The liquid phase is simulated using DNS whereas the dispersed phase is simulated using an IBM with the interaction potential method for deformable interfaces. Our approach extends the IBM of \cite{Spandan+Others.2017} and \cite{Spandan+Others.2018.fast}, where now the dispersed phase is fully coupled both mechanically and thermally to the flow. In addition, two datasets are simulated with and without thermal coupling to investigate its influence on the heat transport. Although $\Nus$ is slightly larger when the droplets are thermally coupled, we found that the VC flow with light droplets exhibits a non-monotonic change in heat transport with increasing $\Ray$ and largely retains the laminar-like VC scaling. We reason that a significant enhancement of heat transport depends crucially on a sufficiently strong droplet driving, which we show can be characterised by the relative strength of $\Ray_d$ to $\Ray$ and the bubblance parameter, $b$. 

When light droplets are introduced, the mean velocity and temperature profiles are highly skewed with the lowest $\Ray$ being most sensitive (figures \ref{fig:MeanStatsInZ} to \ref{fig:TurbStatsInZ}). However, this sensitivity is masked by the $\Nus$ versus $\Ray$ trend, where we observe a non-monotonic behaviour with increasing $\Ray$ (figure \ref{fig:NuRaComp}$a$). This suggests the presence of competing mechanisms in the flow that contribute to the net heat transport. In contrast, the decreasing $\Rey$ versus $\Ray$ trends are commensurate with the higher sensitivity at lower $\Ray$, \ie\,mechanical stirring is strongest at lowest $\Ray$ and higher $\alpha$ (figure \ref{fig:NuRaComp}$b$).

Based on analyses of the near-wall regions, we found that regions with higher values of local heat fluxes, $\Nus_{\text{loc}}$, correspond to concomitant regions with higher values of skin-friction coefficient, $C_{f,\text{loc}}$, which is consistent with the notion that the local wind has influence over the local heat transport (figures \ref{fig:LocalNuTrends} and \ref{fig:LocalCfTrends}). In turn, the strength of the local wind is related to whether the direction of the rising droplets aids or opposes the flow (figure \ref{fig:LocalCfRatio}). However, the trends of $\Nus_{\text{loc}}$ and $C_{f,\text{loc}}$ remain spatially non-monotonic and is sensitive to $\alpha$. Based on these observations, we deduce that the results are specific to our selected simulation parameters and stress that the trends are non-universal.

The $\Nus$ versus $\Ray$ trend in figure \ref{fig:NuRaComp} is different \dl{from} recent experimental results by \cite{Gvozdic+Others.2018} \dl{for bubbly flow}. Whilst $\Nus$ exhibits some $\Ray$-dependency for our simulations with light droplets, \cite{Gvozdic+Others.2018} reported that $\Nus$ is largely insensitive to $\Ray$ for various volume fractions of droplets. The key distinction between our DNS and the experiments by \cite{Gvozdic+Others.2018} becomes readily apparent when we quantify the bubblance parameter $b$ and the droplet driving parameter $\Ray_d/\Ray$ (\cf\,\S\,\ref{sec:ComparisonWithExp}). Both $b$ and $\Ray_d/\Ray$ have a large separation in scales between the laboratory experiments and our DNS. More specifically, at $b \gtrsim O(10^{-1})$ and $\Ray_d/\Ray \gtrsim O(100)$, we anticipate that the dynamics of the dispersed phase-induced liquid agitations become overwhelmed by the dynamics of the background VC flow. For light droplets, both $b$ and $\Ray_d/\Ray$ are significantly lower. \dl{Therefore} the local heat fluxes and skin friction coefficients exhibit non-universal behaviour, which reflects the dominance of the background VC flow.

Our results collectively indicate a non-universal heat transport behaviour for light droplets. Locally, the near-wall trends of heat fluxes and wall-shear stresses suggest the presence of competing mechanisms that, in concert, govern heat transport. One question that arises naturally here \dl{is: Can} a universal trend be eventually obtained by increasing $b$ and $\Ray_d/\Ray$ for fixed $\Ray$? The answer to this question may provide some clues on disentangling the competing heat transport mechanisms in multiphase VC and is a subject for our future investigations.

\section*{Acknowledgements}
We wish to express our gratitude to A.\,Prosperetti for the various fruitful discussions.
This work is part of the research programme of the Foundation for Fundamental
Research on Matter with project number 16DDS001, which is financially supported by the Netherlands Organisation for Scientific Research (NWO). The simulations were carried out on the national e-infrastructure of SURFsara, a subsidiary of SURF cooperation, the collaborative ICT organization for Dutch education and research. We also acknowledge PRACE for awarding us access to MareNostrum hosted by the Barcelona Supercomputing Center (BSC), Spain, under PRACE project number 2017174146. 

\section*{Declaration of Interests}
The authors report no conflict of interest.

\bibliographystyle{jfm}

\end{document}